\documentclass[10pt,english]{article}
\usepackage[T1]{fontenc}
\usepackage[latin9]{inputenc}
\usepackage{geometry}
\usepackage{float}
\usepackage{mathtools}
\usepackage{amsmath}
\usepackage{amssymb}
\usepackage{graphicx}
\usepackage{esint}
\def\be{\begin{equation}}
\def\ee{\end{equation}}

\makeatletter

\providecommand{\tabularnewline}{\\}
\newcommand{\lyxdot}{.}

\makeatother

\usepackage{babel}
\begin{document}

\title{\bf Short-scale Emergence of Classical Geometry,\\  
in Euclidean Loop Quantum Gravity}

\author{Vincent Bayle, Fran\c{c}ois Collet and Carlo Rovelli\\CPT, CNRS Case
907, Aix-Marseille University, 13288 Marseille, France}
\maketitle
\begin{abstract}

\noindent We study the euclidean covariant loop-quantum-gravity vertex numerically, 
using a cylindrically symmetric boundary state and a convenient value of
the Barbero-Immirzi parameter. We show that a classical 
geometry emerges already at low spin.  We also recognise 
the appearance of the degenerate configurations.
\end{abstract}

\section{Introduction}

A celebrated theorem by Barrett \emph{et.al.} \cite{Barrett:2009ij} (see also Conrady and Freidel \cite{Conrady:2008ea}) states that the vertex amplitude \cite{Engle:2007uq,Freidel:2007py,Engle:2007wy,Kaminski:2009fm} of loop quantum gravity admits a geometrical interpretation in terms of the geometry of a four-simplex, its value being determined by the  Regge action of this four-simplex. The theorem has been extended to the Lorentzian theory \cite{Barrett2009}, to the physical case of positive cosmological constant \cite{Han:2011aa,Fairbairn:2010cp}, and is at the basis of a number of results relating the quantum dynamics of loop gravity to classical general relativity \cite{Conrady:2008ea,Magliaro2011,MAGLIARO2013,Han:2013tap,Haggard}, which are at the foundation of the covariant formulation of loop gravity \cite{Rovelli:2014ssa}. All these results are derived in the large spin limit; namely under the assumption that the vertex describes (to low order) a process in a region of spacetime large compared to the Planck size.  Here, we study the behavior of the vertex amplitude for small spins. In particular, we compute the amplitude for some cylindrically symmetric geometries on the boundary of the vertex and we evaluate it numerically. 

In the wake of previous similar results \cite{Magliaro2007,Bianchi:2011ub,Bianchi2012e}, we find that the results proven mathematically in the limit $j\to\infty$ actually hold true already at rather small spin $j$, namely for vertices representing  spacetime regions of Planckian size.  We find evidence for emergence of semiclassical behaviour already for $j\sim 10$, which is to say an order of magnitude above the Planck scale. This might be relevant for instance in cosmology, suggesting that quantum gravitational effects could be limited to to regimes very near Planckian densities. 

On the other hand, we also see the appearance of genuine quantum phenomena in the numerical result.  These are first of all the spread of the amplitude around the classical values, which is simply Heisenberg uncertainly. But also the emergence of degenerate geometries, on which we comment in closure. 

Our analysis has three main limitations.  First, it is in the euclidean domain instead than the physically relevant Lorentzian domain. The reason we have taken this simplification is only because the Lorentzian vertex appears to be algebraically more complicated. The euclidean vertex can be simply expressed in terms of Wigner $n-j$ symbols, which can be directly  handled numerically.  The euclidean theory, on the other hand, has an intrinsic difficulty (absent in the Lorentzian one), which is that for generic values of the Barbero-Immirzi parameter $\gamma$ the simplicity conditions between (discrete) spins cannot be satisfied. We have circumvented this obstacle by choosing $\gamma=1/2$. We do not know how bad this is. Finally, we limit our analysis to (Livine-Speziale \cite{Livine2008a}) boundary states with a convenient ``cylindrical'' symmetry. This choice makes the problem tractable.  Geometrically, this corresponds to studying a 4-simplex whose geometry is invariant under cyclic permutations of three of its five boundary tetrahedra. 

The paper is organised as follows.  We begin by studying the geometry of a 4-simplex. In particular, first, we express this geometry in terms of variables that are the natural variables in quantum gravity: the areas of its 2d triangles (corresponding to the  spins of loop gravity) and suitable variables to capturing the  shape of the tetrahedra  (corresponding to the  intertwiners of loop gravity); second, we find the relation between these implied by the symmetry assumed.   Then we write the quantum amplitude of a coherent state with these symmetries on the boundary of a 4-simplex, and we evaluate the amplitude numerically.  We show that the amplitude is suppressed for the quantum configurations that do not correspond to classical geometries, as in the  Barrett \emph{et.al.} theorem. We summarise our results in the conclusion.  

\section{Geometry and classicality conditions}

The geometrical object we consider is a flat 4-simplex. By this we mean a portion of flat 4d space bounded by a 3d surface $\Sigma$ formed by five flat tetrahedra $\tau_k$, $k=1,...,5$ matching along their boundary triangles $t_{kl}$. The geometry of this object is determined (possibly up to parity) by ten numbers, which can be taken to be the length of its ten segments or the area $a_{kl}$ of its ten triangles. We take the areas $a_{kl}$ of the triangles as basic variables for the geometry, because they play a related role in the quantum theory. 

The geometry of an \emph{individual} tetrahedron $\tau$ is determined by six numbers.  These can be taken to be the length of its six segments, or the four areas $a_i$, with $i=1,...,4$ of its four triangles plus two other variables that capture its \emph{shape} at fixed value of the areas.  A convenient choice is given by the two variables $\Phi$ and $A$ defined below. 

Since the geometry of the 4-simplex is determined by the ensemble of all the areas $a_{kl}$, the value of the shape variables $\Phi_k$ and $A_k$ of a tetrahedron $\tau_k$ sitting on the boundary of a 4-simplex is in fact determined non-locally by the ensemble of all the area $a_{kl}$ around the 4-simplex. (This can be conventionally seen as an effect of the dynamics that glues all tetrahedra together.) Therefore there exist well defined functions $\Phi_k(a_{kl})$ and $A_k(a_{kl})$ giving the shape of each (classical) tetrahedron, as a function of all the ten areas of the 4-simplex.

In quantum gravity, we can associate to each tetrahedron $\tau$ with areas $a_i, i=1,...,4$, a coherent quantum state $|a_i,(\Phi,A)\rangle$ picked on a given shape $(\Phi,A)$.  The boundary of the fundamental vertex of the theory is formed by five such quantum tetrahedra, with matching areas. Therefore on the boundary we can place states determined by the ten areas $a_{kl}$ but peaked on arbitrary shapes: $|a_{kl},(\Phi_k,A_k)\rangle$. The corresponding amplitude is written 
\be
W(a_{kl},\Phi_k,A_k)=\langle W|a_{kl},(\Phi_k,A_k)\rangle
\ee
A main result of the Barrett \emph{et.al.} theorem is that this amplitude is exponentially suppressed for large areas unless $\Phi_k=\Phi(a_{kl})$ and $A_k=A_k(a_{kl})$. That is, the quantum vertex implements these classicality conditions on the boundary state. We are interested to study whether these condition are also implemented at small areas. \\

\subsection{Geometry of a tetrahedron}

Consider a (flat) tetrahedron in 3d flat space. Let the index $i=1,2,3,4$ label its faces and denote $a_i$ the area of the face $i$. Let $\overrightarrow{n_i}$ be a vector of unit length normal to the face $i$. Elementary geometry gives:
\be
\sum_{i=1}^{4}\ a_{i}\overrightarrow{n_{i}}=0
\ee
(Proof: Immerge the tetrahedron in water and pressure the water. Would it move?  Clearly not. Therefore the sum of the forces due to the pressure on its faces, which is the l.h.s.  of this equation, must vanish...)  Now pair arbitrarilly the faces ---say: $(1,2),(3,4)$--- and define the angle $\Phi$ betwen the planes $\left(\overrightarrow{n_{1}},\overrightarrow{n_{2}}\right)$
and $\left(\overrightarrow{n_{3}},\overrightarrow{n_{4}}\right)$
:
\be
\cos\Phi\coloneqq-\left(\frac{\overrightarrow{n_{1}}\wedge\overrightarrow{n_{2}}}{\left\Vert \overrightarrow{n_{1}}\wedge\overrightarrow{n_{2}}\right\Vert }\right)\cdot\left(\frac{\overrightarrow{n_{3}}\wedge\overrightarrow{n_{4}}}{\left\Vert \overrightarrow{n_{3}}\wedge\overrightarrow{n_{4}}\right\Vert }\right)
\ee
and the ``projected area'' $A$ by
\be
 a_{1}\overrightarrow{n_{1}}+a_{2}\overrightarrow{n_{2}}=-\left(a_{3}\overrightarrow{n_{3}}+a_{4}\overrightarrow{n_{4}}\right)\coloneqq A \ \overrightarrow{n_{P}}
\ee
where $\overrightarrow{n_{P}}$ has unit norm.  The six quantities $(a_i,A,\Phi)$ determine the shape of the tetrahedron. 

In particular the normals $\overrightarrow{n_{i}}$ can be computed from these six quantities, choosing an orientation (or ``gauge'') for the tetrahedron. A choice of orientation and the explicit formulae for the the normals as functions of $(a_i,A,\Phi)$ are given in Appendix A.

\subsection{The cylindrically symmetric 4-simplex}\label{Sc}

\begin{figure}
\begin{center}
\includegraphics[scale=0.2]{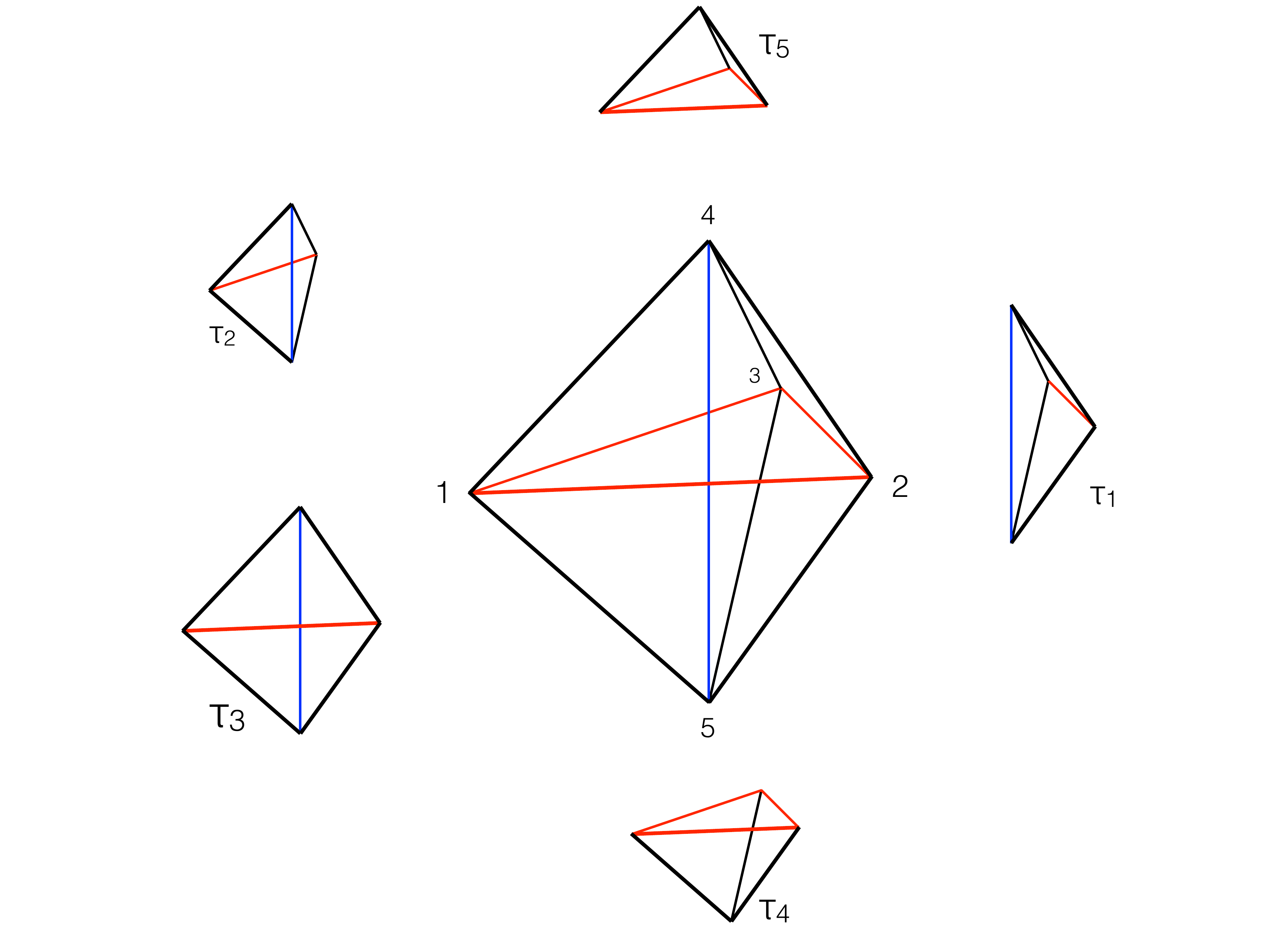}\hspace{3em}
\includegraphics[scale=0.2]{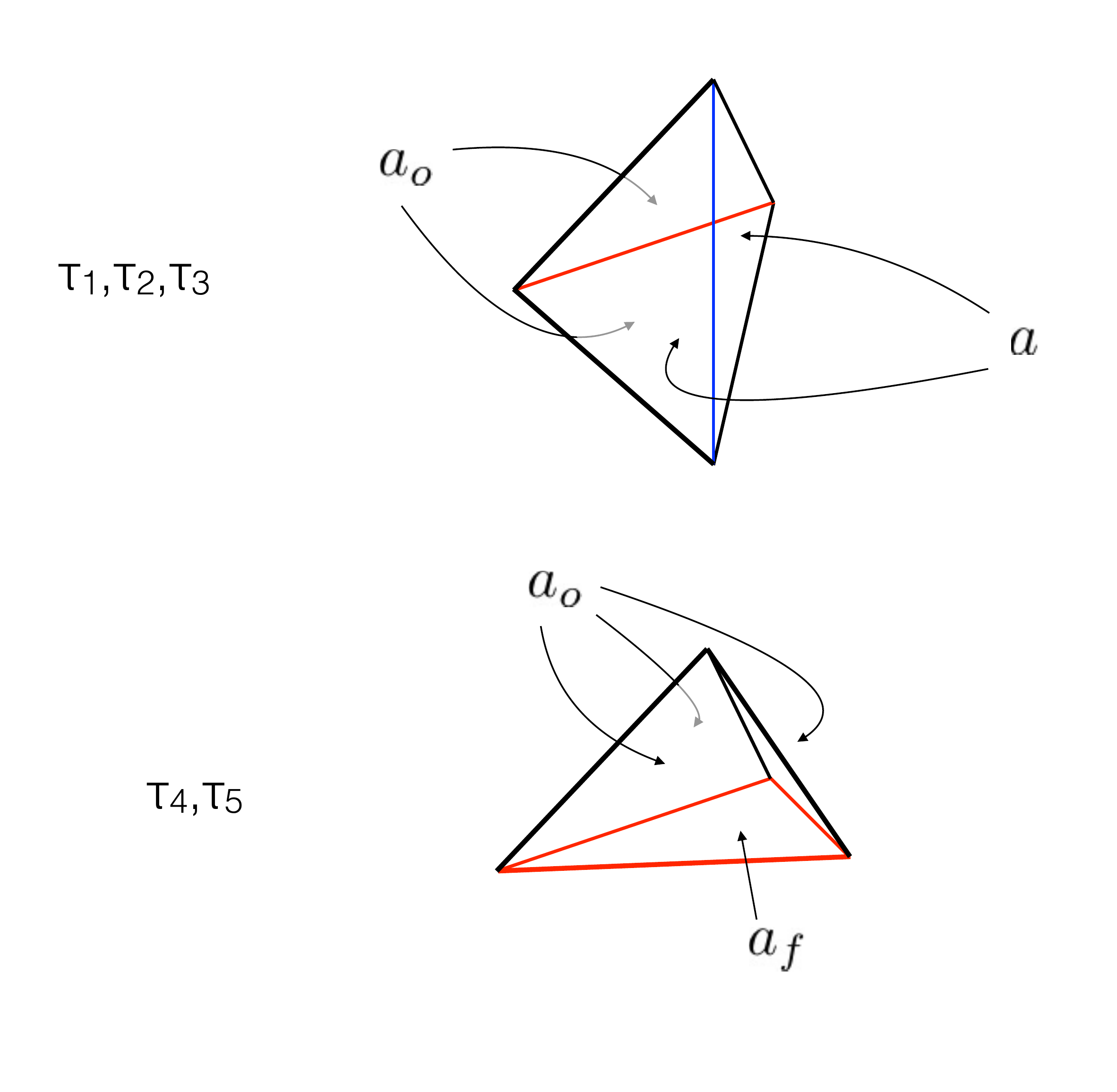}
\caption{Left: The cylindrically symmetric 4-simplex and its boundary tetrahedra, with the areas of their faces.  Right: The areas of the two triangle's shapes; the equatorial ones (up) 
and the polar ones (down).  }
\label{uno}
\end{center}
\end{figure}

In this paper we study the elementary quantum geometry of a regioni of space with cylindrical symmetry.  In the classical case, this is defined by a 4-simplex invariant under cyclic permutation of three of its boundary tetrahedra, say $\tau_1,\tau_2$ and $\tau_3$.  We call these three tetrahedra ``equatorial'' and the other two, namely $\tau_4$ and $\tau_5$ ``polar''.  See Figure 1.  We call $P_k$ the vertex of the 4-simplex opposite to the tetrahedron $\tau_k$ and $L_{kl}$ the length if the segment joining $P_k$ and $P_l$, then the symmetry implies that the geometry is entirely determined by three lengths:
\be
L_{12}=L_{23}=L_{31}, \ \ \ \ \ \ \ \  L_{14}=L_{24}=L_{34}=L_{15}=L_{25}=L_{35} \ \ \ \ \ {\rm and } \ \ \  L_{45},
\ee
which are respectively red, black and blue in Figure 1.  Calling $a_{kl}$ the area of the triangles opposite to the segment $L_{kl}$ (separating the tetrahedra $\tau_k$ and $\tau_l$), we use the notation  
\be
a_{12}=a_{23}=a_{31} \equiv a, \ \ \ \ \ \ \ a_{14}=a_{24}=a_{34}=a_{15}=a_{25}=a_{35} \equiv a_0 \ \ \ \ \ {\rm and } \ \ \ \ \  a_{45}\equiv a_f.
\label{symmetry}
\ee
The three areas $a,a_0,a_f$ determine the geometry (up to parity).  Notice that the three equatorial tetrahedra $\tau_1,\tau_2$ and $\tau_3$ have two isosceles faces with area $a_0$ and two isoscele faces, with area $a$. While the two polar tetrahedra $\tau_4$ and $\tau_5$ have three isosceles faces with area $a_0$ and one equilateral face with area $a_f$. See the right panel of Figure\ref{uno}.  

The symmetry implies that the $\Phi$ angle of all tetrahedra are equal to $\frac\pi2$, while their $A$ variable can be determined from elementary geometry. Writing
\be
A_1=A_2=A_3\equiv A,
\ee
for the equatorial tetrahedra, and 
\be
A_4=A_5\equiv A_f,
\ee
for the polar ones, geometry gives 
\be
A_{f}=\sqrt{a_{0}^{2}+\frac{1}{3}a_{f}^{2}}
\label{c1}
\ee
and
\be
A^{2}\frac{4a_{0}^{2}-A^{2}}{4a^{2}-A^{2}}=\frac{4}{3}a_{f}^{2}.
\label{c2}
\ee
The relation (\ref{c1}) come from the symetry of the polar tetrahedra, and exist independently of the closure condition of the 4-simplex. The relation (\ref{c2}) come from the geometry of the equatorial tetrahedra and the matching condition of the segments with the polar tetrahedra inside the 4-simplex. Solving this equation for $A$ gives two solutions, corresponding to  two possible isometries that preserve the lenghts of all segments. 
The geometrical constraint of the equatorial tetrahedra and the closure condition of the 4-simplex picks one of these, which is 
\be
A=\sqrt{2\left(a_{0}^{2}+\frac{1}{3}a_{f}^{2}\right)-2\sqrt{\left(a_{0}^{2}+\frac{1}{3}a_{f}^{2}\right)^2-\frac{4}{3}a^{2}a_{f}^{2}}}.
\label{c3}
\ee
In Appendix A we give an explicit expression for all the normals in terms of $a, a_0, a_f$ alone. 

In summary, the cylindrically symmetric 4-simplex is bounded by two kinds of tetrahedra: three equatorial tetrahedra with triangle areas $(a,a,a_0,a_0)$ and shape variables $(\frac\pi2,A)$; and two polar tetrahedra with triangles areas $(a_0,a_0,a_0,a_f)$ and shape variables $(\frac\pi2,A_f)$. The two shape variables $A$ and $A_f$ are determined by the areas via equations \eqref{c1} and \eqref{c3}, when this boundary encloses the flat cylindrically symmetric 4-simplex. 

\section{Quantum geometry}

\begin{figure}
\begin{center}
\includegraphics[scale=0.25]{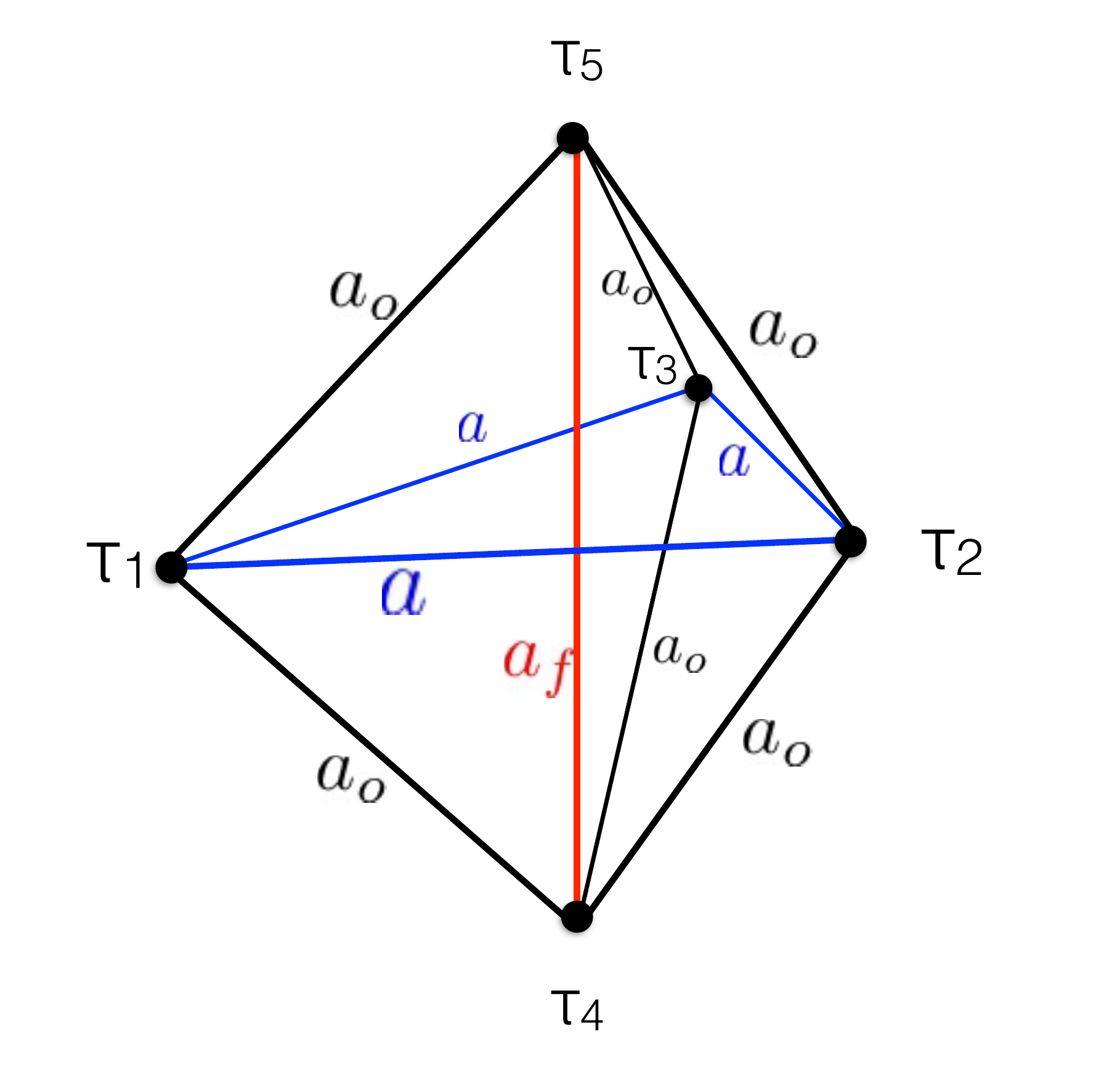}
\hspace{3em}
\includegraphics[scale=0.4]{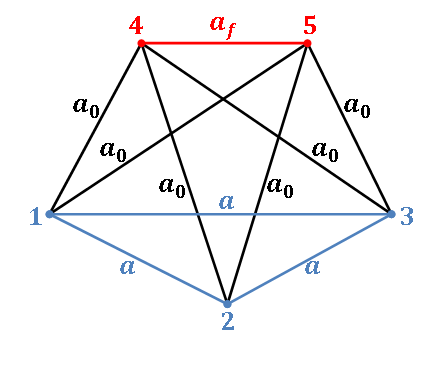}
\caption{Cylindrically data of the boundary spin network, in two equivalent representations.}
\label{due}
\end{center}
\end{figure}

In covariant Loop Gravity, states are defined on the 3d boundary of a spacetime region.  A basis of states is given by the spin network states, that have support on a graph that can be interpreted as the dual of a 3d triangulation.  The theory associates an amplitude to such boundary states. The amplitude can be computed using the spinfoam expansion: at each order the amplitude is given a by a spinfoam defined on a two-complex whose boundary is the graph of the boundary state. In particular, the spinfoam can be defined on the dial of a a triangulation of the spacetime region. Here we consider the lowest order of the expansion, where the 4d triangulation is formed by a single 4-simplex. Figure \ref{due} gives the graph of the boundary triangulation of the 4-simplex. Although formally similar, this represents actually the graph \emph{dual} to that of Figure \ref{uno}: points represent tetrahedra and lines represents triangles. 

Quantum states of the geometry on a boundary are square integral functions $\psi(u_{kl})$ of one $SU(2)$ group variable $u_{kl}$ per each link of the graph. A basis in their space is given by the spin network functions 
\be
\psi^{j_{kl},J_{k}}(u_{kl})=\prod_{\mbox{nodes}\ k}i^{J_{k}}\cdot\prod_{\mbox{links}\ kl}D^{j_{kl}}\left(u_{kl}\right)
\ee
where $j_{kl}$ are spins and $J_{k}$ intertwiners, and the contraction is dictated by the topology of the graph. 

These states are eigenstates of the area of the triangles, with eigenvalues 
\be
a_{kl}=\frac{8\pi\gamma\hbar G}{c^{3}}\sqrt{j_{kl}\left(j_{kl}+1\right)}
\ee
We chose units where ${8\pi\gamma\hbar G}/{c^{3}}=1$ so we do not have to carry over the dimensionfull factor. 

\subsection{Cylindrical symmetric spin networks}\label{eqv}

We begin implementing the cylindrical symmetry by choosing boundary states where, as in \eqref{symmetry},
\be
j_{12}=j_{23}=j_{31} \equiv j, \ \ \ \ \ \ \ j_{14}=j_{24}=j_{34}=j_{15}=j_{25}=j_{35} \equiv j_0 \ \ \ \ \ {\rm and } \ \ \ \ \  j_{45}\equiv j_f.
\ee
The integers or half-integers $j$, $j_{0}$, $j_{f}$ are the quantum equivalent of the areas $a$, $a_{0}$, $a_{f}$.

Let us now come to the intertwiners.  For the intertwiners between four representations $j_1,...,j_4$, we use a basis defined by 
\be
i_{m_{1}m_{2}m_{3}m_{4}}^{J}=\sqrt{2J+1}\sum_{M}\left(-1\right)^{J-M}\left(\begin{array}{ccc}
j_1 & j_2 & J\\
m_{1} & m_{2} & M
\end{array}\right)\left(\begin{array}{ccc}
j_3 & j_4 & J\\
m_{3} & m_{4} & -M
\end{array}\right),
\ee
where the $\left(\begin{array}{ccc}
j_{1} & j_{2} & j_{3}\\
m_{1} & m_{2} & m_{3}
\end{array}\right)$ are the Wigner 3j-symbols defining the 3-valent invariant of $SU(2)$.
In the case of the equatorial tetrahedra we pair the faces with the same area and write
\be
\begin{tabular}{c}
\includegraphics[scale=0.5]{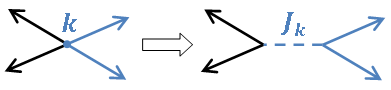}\tabularnewline
$i_{m_{1}m_{2}m_{3}m_{4}}^{J_{k}}=\sqrt{2J_{k}+1}\sum_{M}\left(-1\right)^{J_{k}-M}\left(\begin{array}{ccc}
j & j & J_{k}\\
m_{1} & m_{2} & M
\end{array}\right)\left(\begin{array}{ccc}
j_{0} & j_{0} & J_{k}\\
m_{3} & m_{4} & -M
\end{array}\right)\qquad\left(\mbox{for}\ k=1,2,3\right)$\tabularnewline
\end{tabular}
\ee
While for the polar tetrahedra, we define the matching and the intertwiners as follows 
\be
\begin{tabular}{c}
\includegraphics[scale=0.5]{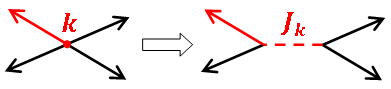}\tabularnewline
$i_{m_{1}m_{2}m_{3}m_{4}}^{J_{k}}=\sqrt{2J_{k}+1}\sum_{M}\left(-1\right)^{J_{k}-M}\left(\begin{array}{ccc}
j_{f} & j_{0} & J_{k}\\
m_{1} & m_{2} & M
\end{array}\right)\left(\begin{array}{ccc}
j_{0} & j_{0} & J_{k}\\
m_{3} & m_{4} & -M
\end{array}\right)\qquad\left(\mbox{for}\ k=4,5\right)$\tabularnewline
\end{tabular}
\ee
The intertwiners $i^{J_{k}}$ associated to a node determines the quantum geometry of the tetrahedron $\tau_k$. The number $J_{k}$, integer or half-integer, is the quantum number of the projected
area $A_{k}$. The following graph illustrates the quantum numbers defining the spin network and the 
chosen pairings for the intertwiners
\be
\includegraphics[scale=0.3]{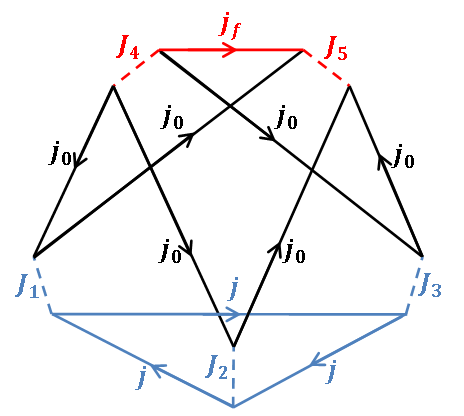}
\ee
Explicitelly, the boundary spin network states we consider are 
\begin{eqnarray}
\psi^{j_{0},j,j_{f},J_{k}}(u_{kl})=&&\!\!\!\!\!\!\!\!\!\!\sum_{m,n}(-1)^{\sum(j_{kl}-n_{kl})}i_{-n_{12}m_{31}-n_{14}m_{51}}^{J_{1}}i_{-n_{23}m_{12}-n_{24}m_{52}}^{J_{2}}i_{-n_{31}m_{23}-n_{34}m_{53}}^{J_{3}}i_{m_{54}m_{34}m_{24}m_{14}}^{J_{4}}i_{-n_{54}-n_{51}-n_{52}-n_{53}}^{J_{5}}
\nonumber \\ && \times \ \ \ 
D_{m_{14}n_{14}}^{j_{0}}\left(u_{14}\right)D_{m_{24}n_{24}}^{j_{0}}\left(u_{24}\right)D_{m_{34}n_{34}}^{j_{0}}\left(u_{34}\right)D_{m_{51}n_{51}}^{j_{0}}\left(u_{51}\right)D_{m_{52}n_{52}}^{j_{0}}\left(u_{52}\right)D_{m_{53}n_{53}}^{j_{0}}\left(u_{53}\right)
\nonumber \\ &&\times \ \ \ 
D_{m_{12}n_{12}}^{j}\left(u_{12}\right)D_{m_{23}n_{23}}^{j}\left(u_{23}\right)D_{m_{31}n_{31}}^{j}\left(u_{31}\right)D_{m_{54}n_{54}}^{j_{f}}\left(u_{54}\right)
\end{eqnarray}
These states are eigenstates of the area operators of the boundary :
\be
\hat a_{kl}\left|\psi^{j_{0},j,j_{f},J_{k}}\right\rangle =\sqrt{j_{kl}\left(j_{kl}+1\right)}\left|\psi^{j_{0},j,j_{f},J_{k}}\right\rangle =\begin{cases}
\sqrt{j\left(j+1\right)}\left|\psi^{j_{0},j,j_{f},J_{k}}\right\rangle  & \mbox{for \ensuremath{k,l=1,2,3}}\\
\sqrt{j_{f}\left(j_{f}+1\right)}\left|\psi^{j_{0},j,j_{f},J_{k}}\right\rangle  & \mbox{for \ensuremath{kl=45}}\\
\sqrt{j_{0}\left(j_{0}+1\right)}\left|\psi^{j_{0},j,j_{f},J_{k}}\right\rangle  & \mbox{else}
\end{cases}
\ee
and satisfy the orthogonality relation :
\be
\left\langle \psi^{j_{0}',j',j_{f}',J_{k}'}|\psi^{j_{0},j,j_{f},J_{k}}\right\rangle =\int_{SU(2)}du\overline{\psi^{j_{0}',j',j_{f}',J'}(u_{kl})}\psi^{j_{0},j,j_{f},J}(u_{kl})=\frac{\delta_{j_{0},j_{0}'}\delta_{j,j'}\delta_{j_{f},j_{f}'}}{\left(2j_{0}+1\right)^{6}\left(2j+1\right)^{3}\left(2j_{f}+1\right)}\prod_{k}\delta_{J_{k},J_{k}'}
\ee

\subsection{Coherent symetric 4-simplex}

The spin network states defined in the previous section are eigenstates of the projected area $A_k$ of the tetrahedra, and are therefore completely spread in the corresponding angles $\Phi_k$, which do not commute with $A_k$.  Therefore they are very non-classical. We are interested, instead, in wave packets that are minimaly spread \emph{both} in $A_k$ and in $\Phi_k$.   To this aim, we use the (intrinsic) coherent states defined by Livine and Speziale \cite{Livine:2006it}.  These are defined as follows.  The coherent link states are defined by 
\be
\left|j\overrightarrow{n}\right\rangle =R\left(\overrightarrow{n}\right)\left|j,j\right\rangle =\sum_{m}D_{mj}^{j}\left(R\left(\overrightarrow{n}\right)\right)\left|j,m\right\rangle 
\ee
where $\overrightarrow{n}$ is the normal vector to a face of tetrahedron with area $j$. The group element $R\left(\overrightarrow{n}\right)$ is 
a rotation than maps the vector $\overrightarrow{u_{z}}$
into the normal vector $\overrightarrow{n}$ :
\be
R\left(\overrightarrow{n}\right)\cdot\overrightarrow{u_{z}}=\overrightarrow{n}
\ee
For a tetrahedron with vectors $\overrightarrow{n_{i}}$ associated to its faces, 
the Livine-Speziale state is:
\be
\left|j_{i}\overrightarrow{n_{i}}\right\rangle =\sum_{m}D_{m_{1}j_{1}}^{j_{1}}\left(R\left(\overrightarrow{n_{1}}\right)\right)D_{m_{2}j_{2}}^{j_{2}}\left(R\left(\overrightarrow{n_{2}}\right)\right)D_{m_{3}j_{3}}^{j_{3}}\left(R\left(\overrightarrow{n_{3}}\right)\right)D_{m_{4}j_{4}}^{j_{4}}\left(R\left(\overrightarrow{n_{4}}\right)\right)\left|j_{1},m_{1}\right\rangle \otimes\left|j_{2},m_{2}\right\rangle \otimes\left|j_{3},m_{3}\right\rangle \otimes\left|j_{4},m_{4}\right\rangle 
\ee
And the projection of this state on the corresponding intertwiner
gives:
\be
\left\langle i^{J}|j_{i}\overrightarrow{n_{i}}\right\rangle =\sum_{m}i_{m_{1}m_{2}m_{3}m_{4}}^{J}D_{m_{1}j_{1}}^{j_{1}}\left(R\left(\overrightarrow{n_{1}}\right)\right)D_{m_{2}j_{2}}^{j_{2}}\left(R\left(\overrightarrow{n_{2}}\right)\right)D_{m_{3}j_{3}}^{j_{3}}\left(R\left(\overrightarrow{n_{3}}\right)\right)D_{m_{4}j_{4}}^{j_{4}}\left(R\left(\overrightarrow{n_{4}}\right)\right)
\ee
Writing $\overrightarrow{n}=\left(\cos\phi\sin\theta,\sin\phi\sin\theta,\cos\theta\right)$, we have 
\be
R\left(\overrightarrow{n}\right)=e^{-\imath\phi J_{Z}}e^{-\imath\theta J_{Y}}
\ee
Where $J_{Z}$ and $J_{Y}$ are the generators of rotations. With this choice of $R$, we can express the $j$-representation :
\be
D_{mj}^{j}\left(R\left(\overrightarrow{n}\right)\right)=D_{mj}^{j}\left(e^{-\imath\phi J_{Z}}e^{-\imath\theta J_{Y}}\right)=e^{-\imath m\phi}d_{mj}^{j}(\theta)
\ee
\be
=\sqrt{\frac{(2j)!}{(j+m)!(j-m)!}}\cdot\frac{\xi^{j-m}}{\left(1+\left|\xi\right|^{2}\right)^{j}}\cdot e^{-\imath j\phi}\qquad\xi=\tan\left(\frac{\theta}{2}\right)e^{\imath\phi}
\ee
Where the $d$ are the little Wigner matrices. The expression of Livine-Speziale
state with his intertwiner became :
\be
\left\langle i^{J}|j_{i}\overrightarrow{n_{i}}\right\rangle =\left(\prod_{i}\frac{e^{-\imath j_{i}\phi_{i}}\sqrt{(2j_{i})!}}{\left(1+\left|\xi_{i}\right|^{2}\right)^{j_{i}}}\right)\sum_{m}i_{m_{1}m_{2}m_{3}m_{4}}^{J}\prod_{i'}\frac{\left.\xi_{i'}\right.^{j_{i'}-m_{i'}}}{\sqrt{(j_{i'}+m_{i'})!(j_{i'}-m_{i'})!}}. 
\ee

In the cylindrical symmetric setting, we have two types of Livine-Speziale distributions:
\be
\left\langle i^{J_{k}}|j,j_{0},A_k,\Phi_k\right\rangle =\left\langle i^{J_{k}}|j_{i}\overrightarrow{n_{i}}\left(A,\Phi\right)\right\rangle \qquad\mbox{for the equatorial tetrahedra \ensuremath{\left(k=1,2,3\right)}}
\ee
\be
\left\langle i^{J_{k}}|j_{f},j_{0},A_k,\Phi_{k}\right\rangle =\left\langle i^{J_{k}}|j_{i}\overrightarrow{n_{i}}\left(A_{f},\Phi_{f}\right)\right\rangle \qquad\mbox{for the polar tetrahedra \ensuremath{\left(k=4,5\right)}}
\ee
These are peaked around the classical geometry define by the variables $j, A,\Phi$ of each tetrahedron. Approximately: 
\be
\left\langle i^{J}|j_{i}\overrightarrow{n_{i}}\left(A,\Phi\right)\right\rangle \propto e^{iJ\Phi}e^{-\frac{\left(J-A\right)^{2}}{2\sigma^{2}(j_{i},A)}}
\ee
These states 
\be
\psi^{j_{0},j,j_{f},A,\Phi,A_f,\Phi_f}(u_{kl})=\sum_{J_{k}}
\left\langle i^{J_{k}}|j_{f},j_{0},A_{k},\Phi_{k}\right\rangle 
\psi^{j_{0},j,j_{f},J_{k}}(u_{kl})
\ee
approximate the intrinsic classical geometry.

\subsection{Spinfoam amplitude}

\begin{figure}
\centerline{\includegraphics[scale=0.15]{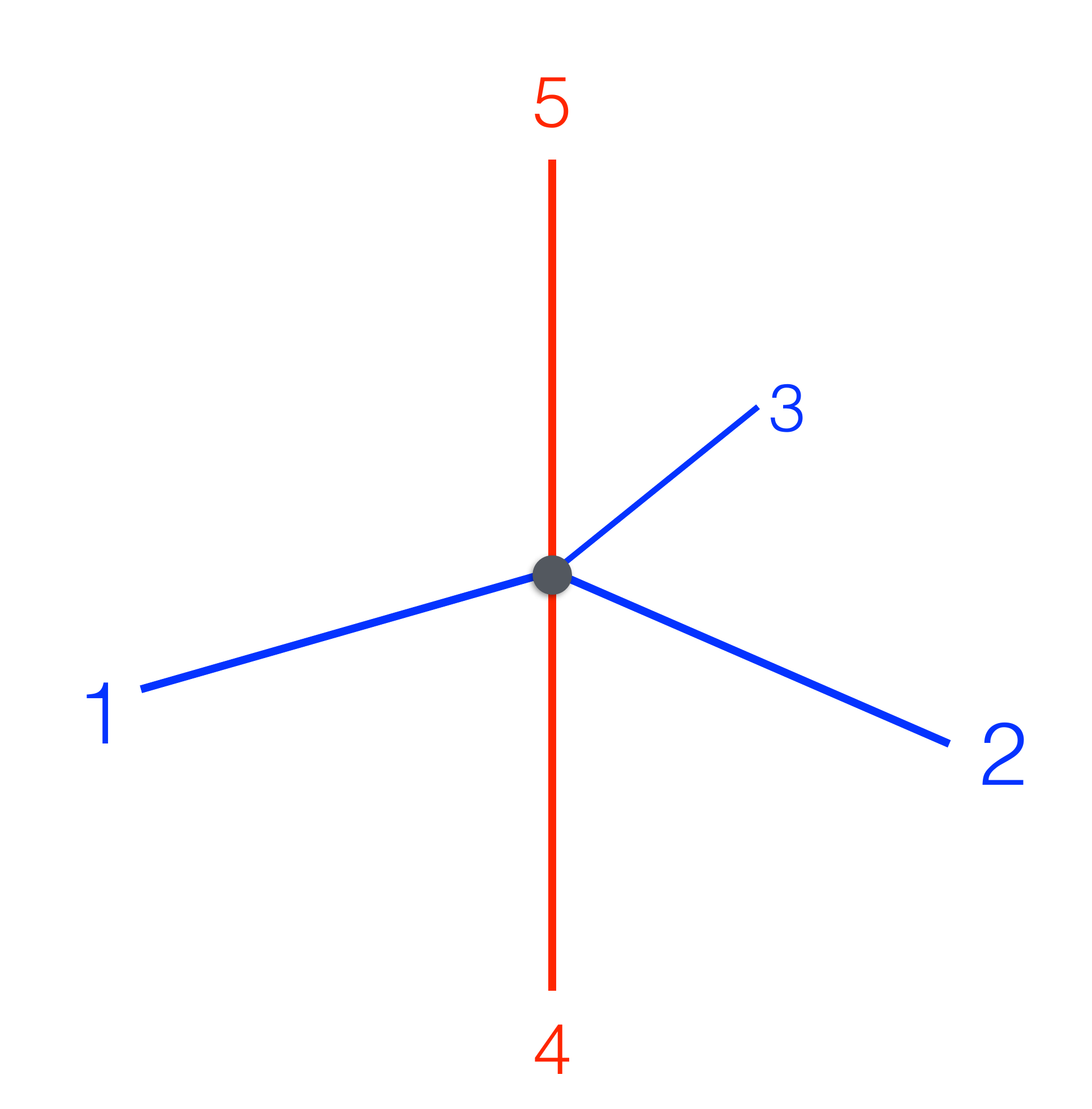}}
\caption{Vertex and edges of the spinfoam.}
\label{quattro}
\end{figure}

We now construct the amplitude associated to the boundary state constructed above. This is given by a single vertex, five edges (See Figure \ref{quattro}) and ten faces. 
The red edges correspond to the polar tetrahedra and the blue edges
are the equatorial tetrahedra; they are connected at single the 4-simplex
vertex.

The covariant LQG amplitude is a function of an $SU(2)$ group element $u_{kl}$ per each face.
It is defined as an integral over 5 copies of $SO(4)\simeq SU(2)^{+}\times SU(2)^{-}$ as follows:
\be
W(u_{kl})=\int_{(SO(4))^5}dU_k\prod_{kl}\delta\left(Y^{+}U_{k}U_{l}^{-1}Yu_{lk}\right)
\ee
where $\delta$ is the $SU(2)$ delta function :
\be
\delta\left(\bullet\right)=\sum_{j}\left(2j+1\right)Tr\left[D^{j}(\bullet)\right]
\ee
and $Y$ is the mapp between the $SO(4)\simeq SU(2)^{+}\times SU(2)^{-}$
bulk variables and the $SU(2)$ boundary variables defined by
\be
Y\ :\ \left|j,m\right\rangle =\sqrt{2j+1}\sum_{m^{+},m^{-}}\left(\begin{array}{ccc}
j^{+} & j^{-} & j\\
m^{+} & m^{-} & m
\end{array}\right)\left|j^{+},m^{+}\right\rangle \otimes\left|j^{-},m^{-}\right\rangle 
\ee
where
\be
j^{\pm}=\frac{1}{2}\left(1\pm\gamma\right)j.
\ee
and $\gamma$ is the Immirzi parameter, which, as mentioned in the introduction, we take to be  $\gamma=\frac{1}{2}$. 

Explicitly, writing $U_{\in SO(4)}=u^{+}\otimes u^{-} \in SU(2)^{+} \otimes SU(2)^{-}$,  this gives
\begin{eqnarray}
Tr\left[D^{j}(Y^{+}UYu^{-1})\right]&=&Tr\left[D^{j}(Y^{+}\left(u^{+}\otimes u^{-}\right)Yu^{-1})\right]
 \\ \nonumber & = &
(2j+1)\sum_{m^\pm, n^\pm}\left(\begin{array}{ccc}
j^{+} & j^{-} & j\\
m^{+} & m^{-} & m
\end{array}\right)\left(\begin{array}{ccc}
j^{+} & j^{-} & j\\
n^{+} & n^{-} & n
\end{array}\right)D_{m^{+}n^{+}}^{j^{+}}\left(u^{+}\right)D_{m^{-}n^{-}}^{j^{-}}\left(u^{-}\right)\overline{D_{mn}^{j}(u)}.
\end{eqnarray}
These definitions give the transition amplitude 
\begin{eqnarray}
W(u_{kl})&=&\sum_{j}\int_{SU(2)^{+}}du_k^{+}\int_{SU(2)^{-}}du_l^{-}\ \prod_{kl}\left(2j_{kl}+1\right)^{2}\\ \nonumber \hspace{-2em}&&
\sum_{m_{kl}^\pm n_{kl}^\pm}
\left(\begin{array}{ccc}
j_{kl}^{+} & j_{kl}^{-} & j_{kl}\\ 
m_{kl}^{+} & m_{kl}^{-} & m_{kl}
\end{array}\right)\left(\begin{array}{ccc}
j_{kl}^{+} & j_{kl}^{-} & j_{kl}\\
n_{kl}^{+} & n_{kl}^{-} & n_{kl}
\end{array}\right)D_{m_{kl}^{+}n_{kl}^{+}}^{j_{kl}^{+}}\left(u_{k}^{+}\left(u_{l}^{+}\right)^{-1}\right)D_{m_{kl}^{-}n_{kl}^{-}}^{j_{kl}^{-}}\left(u_{k}^{-}\left(u_{l}^{-}\right)^{-1}\right)\overline{D_{m_{kl}n_{kl}}^{j_{kl}}\left(u_{kl}\right)}
\end{eqnarray}
The two $SU(2)$ integrals can be performed, giving intertwiners
$i^{K^{+}}$ and $i^{K^{-}}$. geometrically, these correspond to the quantum tetrahedra
in the 4-dimentional euclidean space. The transition amplitude becomes
\be
W(u_{kl})=\sum_{j}\left(\prod_{kl}\left(2j_{kl}+1\right)^{2}\right)\\
\sum_{K^{+},K^{-},K}
\left(K_{k}^{+},j_{kl}^+\right)
\left(K_{k}^{-},j_{kl}^-\right)
\left(\prod_{k}\mathcal{I}_{K_{k}^{+},K_{k}^{-}}^{K_{k}}\left(j_{kl}\right)\right)\overline{\psi^{j_{kl},K_{k}}(u_{kl})}
\ee
Where we have defined the $SU(2)$ 15j-symbol
\be
\left(K_{k},j_{kl}\right)
=\sum_{p}(-1)^{\sum_{kl}\left(j_{kl}-p_{kl}\right)}i_{-p_{12}p_{13}-p_{14}p_{15}}^{K_{1}}i_{-p_{23}p_{12}-p_{24}p_{25}}^{K_{2}}i_{-p_{13}p_{23}-p_{34}p_{35}}^{K_{3}}i_{p_{45}p_{34}p_{24}p_{14}}^{K_{4}}i_{-p_{45}-p_{15}-p_{25}-p_{35}}^{K_{5}}\label{eq:15j}
\ee
And the fusion coefficients 
\be
\mathcal{I}_{K^{+},K^{-}}^{K}\left(j_{a}\right)=i_{m_{1}m_{2}m_{3}m_{4}}^{K}(j_{a})i_{m_{1}^{+}m_{2}^{+}m_{3}^{+}m_{4}^{+}}^{K^{+}}(j_{a}^{+})i_{m_{1}^{-}m_{2}^{-}m_{3}^{-}m_{4}^{-}}^{K^{-}}(j_{a}^{-})\prod_{a=1}^{4}\sqrt{2j_{a}+1}\left(\begin{array}{ccc}
j_{a}^{+} & j_{a}^{-} & j_{a}\\
m_{a}^{+} & m_{a}^{-} & m_{a}
\end{array}\right)
\ee
More compactly, we can define an $SO(4)$ 15j-symbol by 
\be
\left[K_k,j_{kl}\right]
=\sum_{K^{\pm}}
\left(K^+_{k},j^+_{kl}\right)
\left(K^-_{k},j^-_{kl}\right)
\prod_k  \mathcal{I}_{K_{k}^{+}K_{k}^{-}}^{K_{k}}\left(j_{kl}\right)
\ee
where the notation means
 \be
 \begin{aligned}
\mathcal{I}_{K_{1}^{+}K_{1}^{-}}^{K_{1}}\left(j_{1l}\right) & = & \mathcal{I}_{K_{1}^{+}K_{1}^{-}}^{K_{1}}\left(j_{12},j_{13},j_{14},j_{15}\right)\\
\mathcal{I}_{K_{2}^{+}K_{2}^{-}}^{K_{2}}\left(j_{2l}\right) & = & \mathcal{I}_{K_{2}^{+}K_{2}^{-}}^{K_{2}}\left(j_{23},j_{12},j_{24},j_{25}\right)\\
 & \vdots\\
\mathcal{I}_{K_{5}^{+}K_{5}^{-}}^{K_{5}}\left(j_{5l}\right) & = & \mathcal{I}_{K_{5}^{+}K_{5}^{-}}^{K_{5}}\left(j_{45},j_{15},j_{25},j_{35}\right)
 \end{aligned} 
 \ee
like the indices conventions in the spin network section (\ref{eqv}) and in the 15j-symbol \eqref{eq:15j}.

With this, the amplitude of a boundary spin network state is 
\be
W(u_{kl})=\sum_{j}\left(\prod_{kl}\left(2j_{kl}+1\right)^{2}\right)\sum_{K}
\ \left[K_k,k_{kl}\right]\ \overline{\psi^{j_{kl},K_{k}}(u_{kl})}
\ee
The amplitude of a boundary spin network state is therefore simply given by 
\be
W^{j_{kl},J_{k}}=\left\langle W|\psi^{j_{kl},J_{k}}\right\rangle =\int_{SU(2)}du_{kl}\ W(u_{kl})\ \psi^{j_{kl},J_{k}}(u_{kl})
=\left(2j_{0}+1\right)^{6}\left(2j+1\right)^{3}\left(2j_{f}+1\right)\ \left[J_k,j_{kl}\right]
\ee

Finally, we write the amplitude for a coherent cylindrically symmetric boundary state. This, we recall, is determined by three spins, $j,j_0$ and $j_f$ and by the shape variables $A_k,\Phi_k$ of the five tetrahedra.  From the definition of the coherent cylindrically symmetric state in the previous Section, we have immediately 
\be
W(j_{0},j,j_{f},A_k,\Phi_k)=\sum_{J_k} \left\langle W|\psi^{j_{0},j,j_{f},J_{k}}\right\rangle \prod_{k}\left\langle i^{J_{k}}|j,j_{0},A,\Phi\right\rangle.
\ee

This completes the derivation of the amplitude.  In the next Section, we study $W(j_{0},j,j_{f},A_k,\Phi_k)$ numerically. 

\section{Numerical analysis of amplitude}

Barret \emph{et al.}'s theorem  \cite{Barrett:2009ij} states that the vertex amplitude for a coherent boundary state is exponentially suppressed in the large spin limit ($j_{kl}\gg 1$) unless the shapes of the boundary tetrahedra are those determined non-locally by the classical flat geometry of 4-simplex, in terms of the areas of the faces, namely by the $j_{kl}$ themselves.  In the case we are considering, this means that the shape variables $A_k,\Phi_k$ must take the ``classical values'', functions of $j,j_0,j_f$ for the amplitude not to be suppressed. 

We have studied these classical values in  Section \ref{Sc}.  For the angles, they are $\Phi_k=\frac\pi2$ for all $k$'s. For the $A_k$ variables, they are given by the functions  $A_k(j,j_0,j_f)$ defined by the constraint \eqref{c3} for $k=1,2,3$ and by the constraint \eqref{c1} for $k=4,5$ (in the sense of the areas $(a,a_0,a_f)\sim(j,j_0,j_f)$ in the spin network state, see Section \ref{eqv}).  Thus, fixing \emph{large} values of the spins $j,j_0,j_f$, we expect the amplitude $W(j_{0},j,j_{f},A_k,\Phi_k)$, seen as a function of the $A_k$ and the $\Phi_k$ to be peaked on the classical values $\Phi_k=\frac\pi2$ and $A_k=A_k(j,j_0,j_f)$.  We are interested to explore what happens for \emph{small} spins. 

To this aim, we have designed a c++ program that computes the amplitude $W(j_{0},j,j_{f},A_k,\Phi_k)$.  Ideally, we would like to fix the spins and study the peakedness properties of the real function of ten variables 
\be
f_{j_{0},j,j_{f}}(A_k,\Phi_k)=|W(j_{0},j,j_{f},A_k,\Phi_k)|
\ee
However, the ten dimensional space $A_k,\Phi_k$ is too large to explore numerically. So, we study it gradually by exploring some of its sections. 

\subsection{Sections of the space of shapes}

\begin{figure}[H]
\centerline{\includegraphics[scale=0.3]{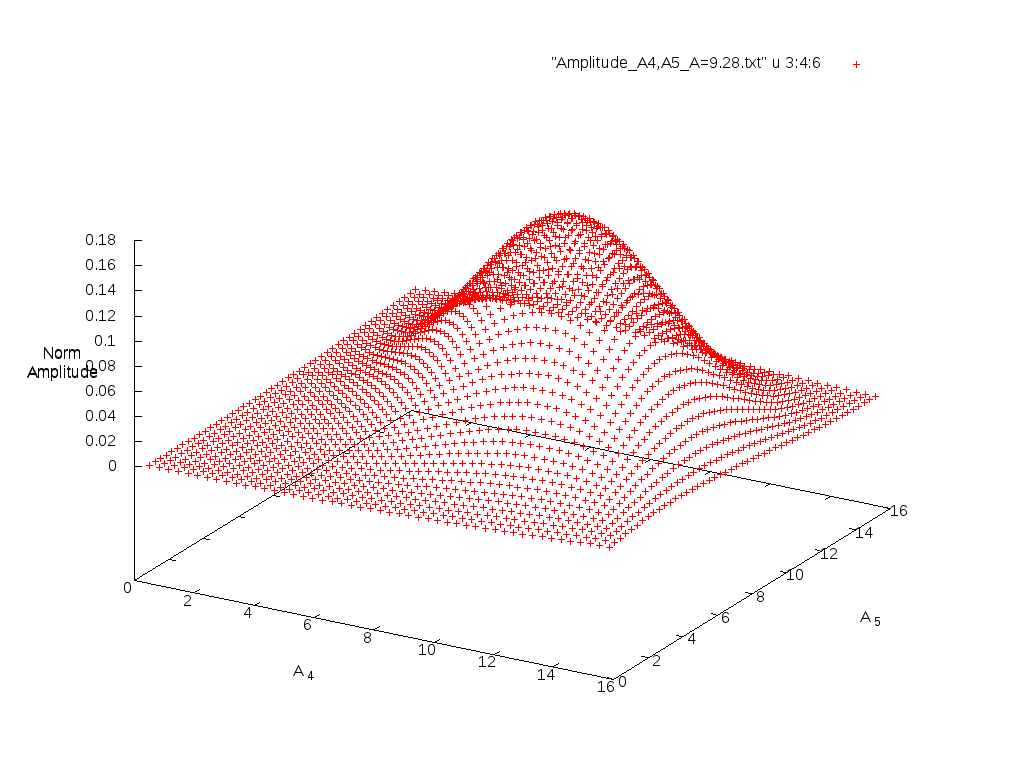} \hspace{2em} \includegraphics[scale=0.3]{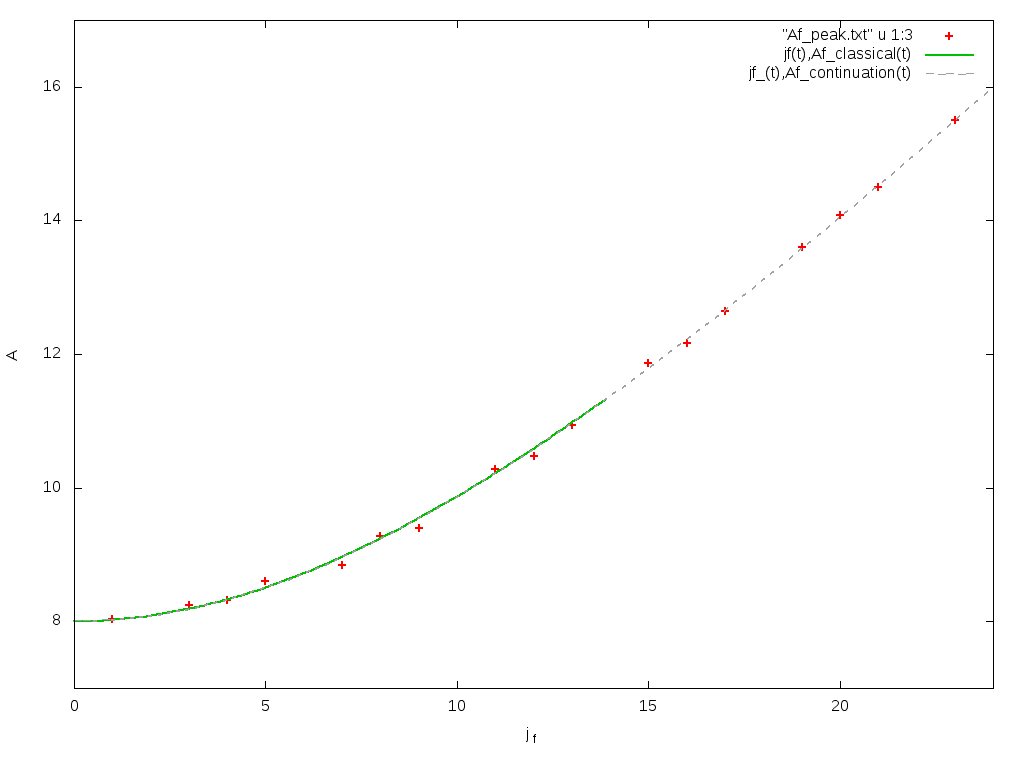}}
\caption{Left: $f(A_4,A_5)$ for $j=j_{0}=j_{f}=8$. Right: The position of the peak as $j_f$ varies (crosses), compared with the classical value (line) and the analytic continuation of the classical value (dotted line).}
\label{FG1}
\end{figure}

To start with, we fix all the angles and all the equatorial projected areas to their classical values given respectively by $\Phi_k=\frac\pi2$ and by equation \eqref{c1}. This defines a function of two variables, the projected areas of the two polar tetrahedra
\be
f(A_4,A_5)=f_{j_{0},j,j_{f}}\left(A(j_{0},j,j_{f}),A(j_{0},j,j_{f}),A(j_{0},j,j_{f}),A_4,A_5,\frac\pi2,\frac\pi2,\frac\pi2,\frac\pi2,\frac\pi2\right). 
\ee
A typical numerical result of the numerical calculation is given in the left panel of Figure \ref{FG1}, where this function is plotted for $j=j_{0}=j_{f}=8$. 

The amplitude clearly peaks on a value of $A_4=A_5=A_f$, which is easily recognised precisely on the classical value $A_f=A_f(j_0,j,j_f)$.  We can track the position of this peak as we change $j_f$ and compare it with the classical value of $A_f$ (or its analytic continuation when the triangular conditions are not respected). The result of this numerical analysis is given in the right panel of Figure \ref{FG1}, which shows that the peaks of the amplitude computed numerically (crosses) follow the classical value.  This shows that, quite remarkably, the peakedness properties on the classical values already appears at small spins $j\sim 10$. This pattern is quite general. 

\begin{figure}[H]
\centerline{\includegraphics[scale=0.3]{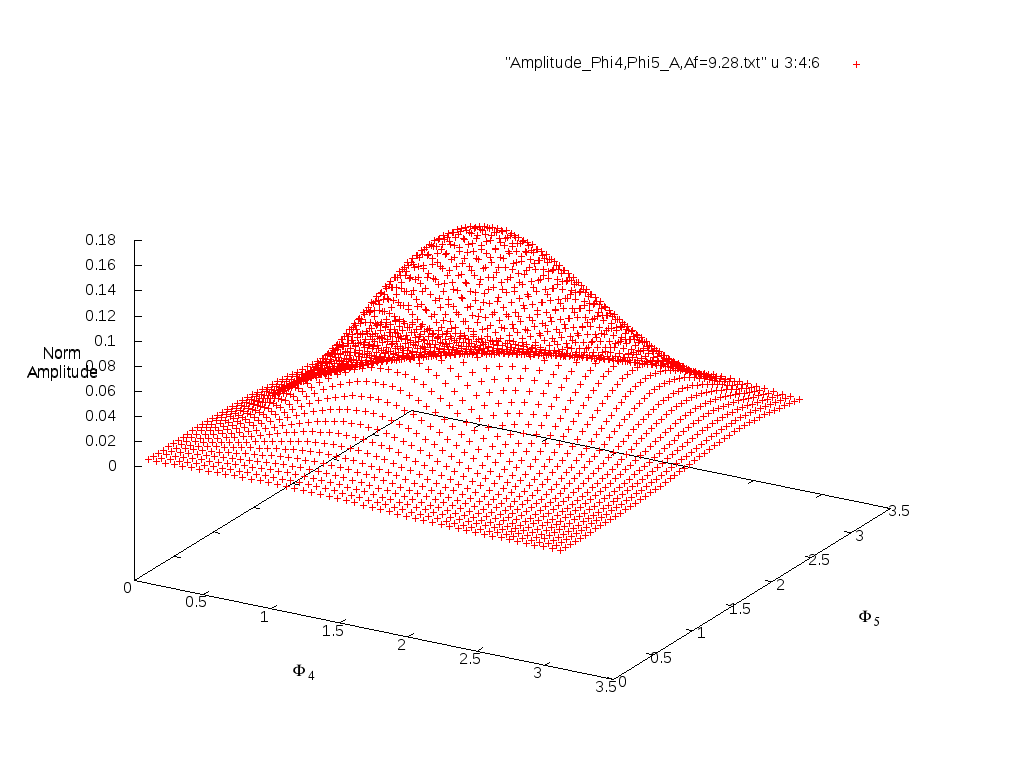}  \hspace{5em} \includegraphics[scale=0.3]{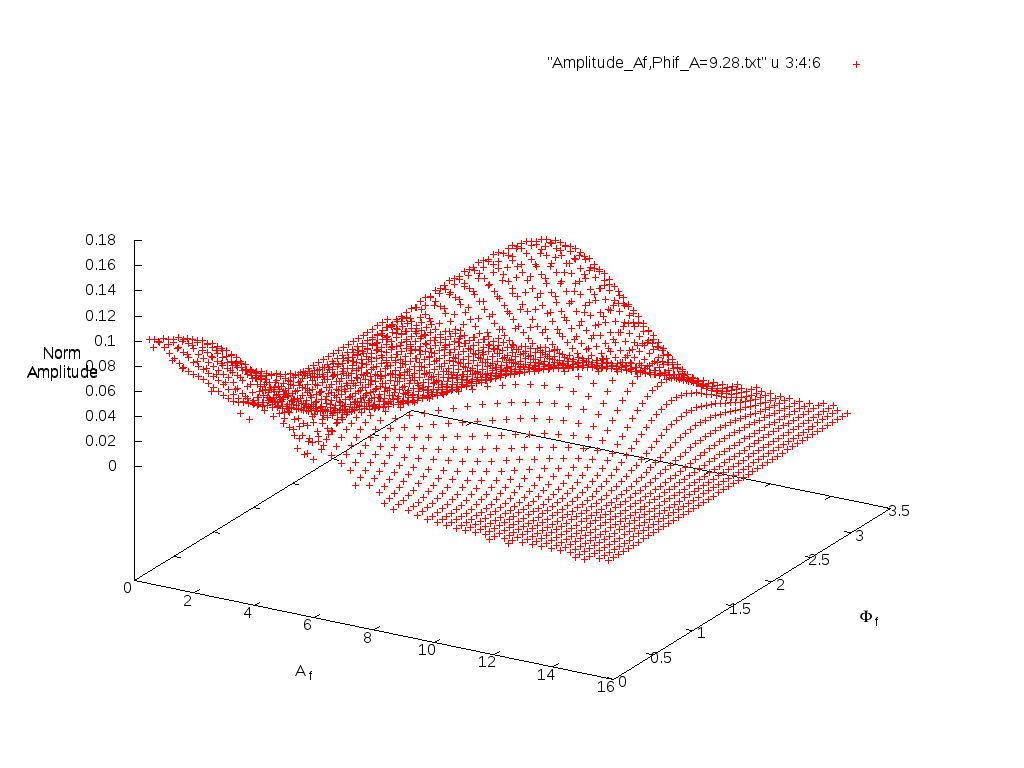}}
\caption{$f(\Phi_4,\Phi_5)$ (Left) and $f(A_f,\Phi_f)$ (Right), for $j=j_{0}=j_{f}=8$.} 
\label{FG2}
\end{figure}

Next, we can reverse the role of the $A$'s and the $\Phi$'s.  That is, we fix all the $A$'s to their classical value and we compute the amplitude as a function of $\Phi_5$ and $\Phi_5$. That is
\be
f(\Phi_4,\Phi_5)=f_{j_{0},j,j_{f}}\left(A(j_{0},j,j_{f}),A(j_{0},j,j_{f}),A(j_{0},j,j_{f}),A_f(j_{0},j,j_{f}),A_f(j_{0},j,j_{f}),\frac\pi2,\frac\pi2,\frac\pi2,\Phi_4,\Phi_5\right). 
\ee
The numerical result is given in the left panel of Figure \ref{FG2}.  We also give the transverse section defined by 
\be
f(A_f,\Phi_f)=f_{j_{0},j,j_{f}}\left(A(j_{0},j,j_{f}),A(j_{0},j,j_{f}),A(j_{0},j,j_{f}),A_f,A_f,\frac\pi2,\frac\pi2,\frac\pi2,\Phi_f,\Phi_f\right). 
\ee
The corresponding numerical result is given in the right panel of Figure \ref{FG2}.  Again we see the peak of the amplitude on the classical values.

\begin{figure}[H]
\centerline{\begin{tabular}{|c|c|c|}
\hline 
$A=0$ & $A=4.16$ & $A=8$\tabularnewline
\hline 
\hline 
\includegraphics[scale=0.2]{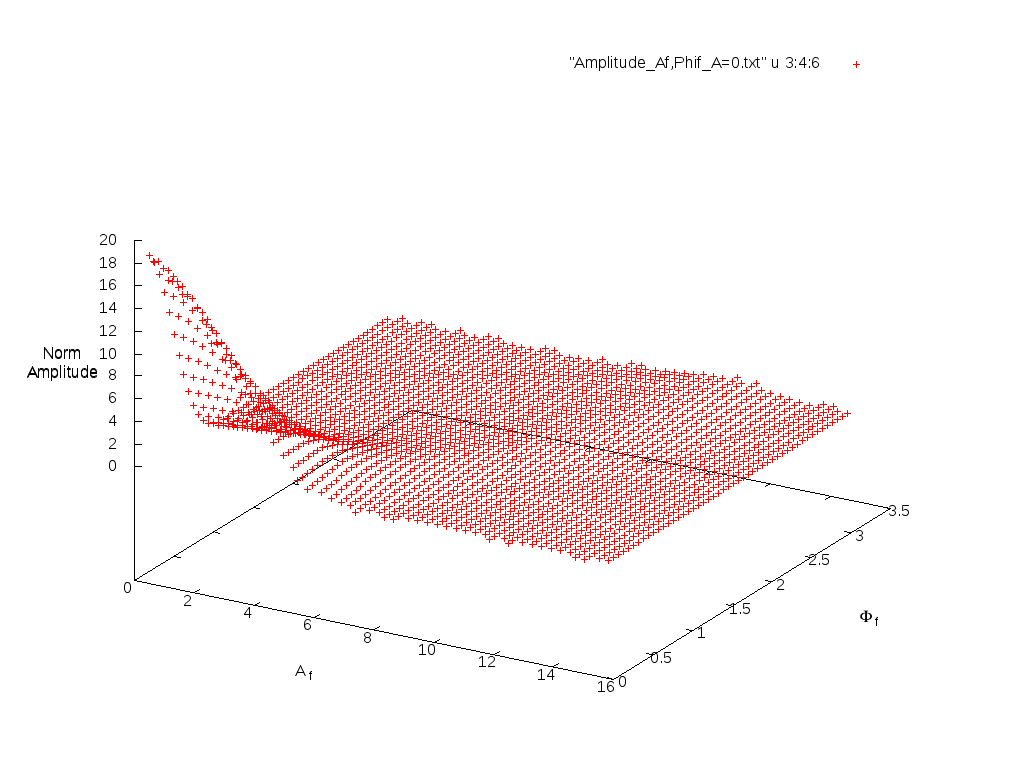} & \includegraphics[scale=0.2]{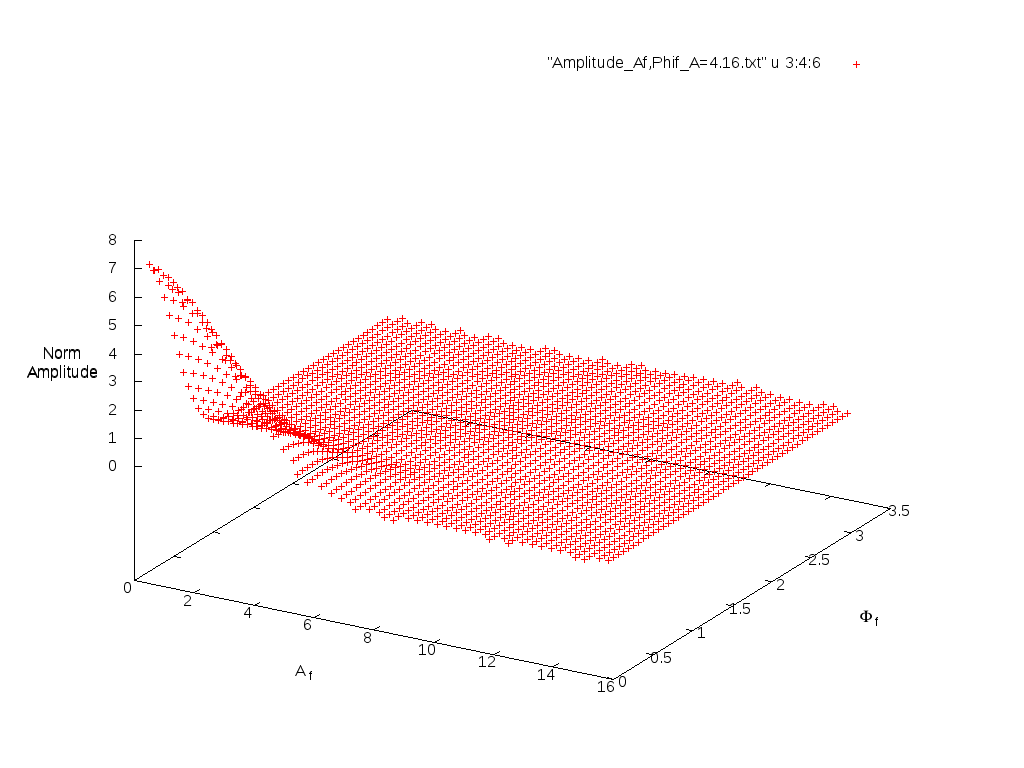} & \includegraphics[scale=0.2]{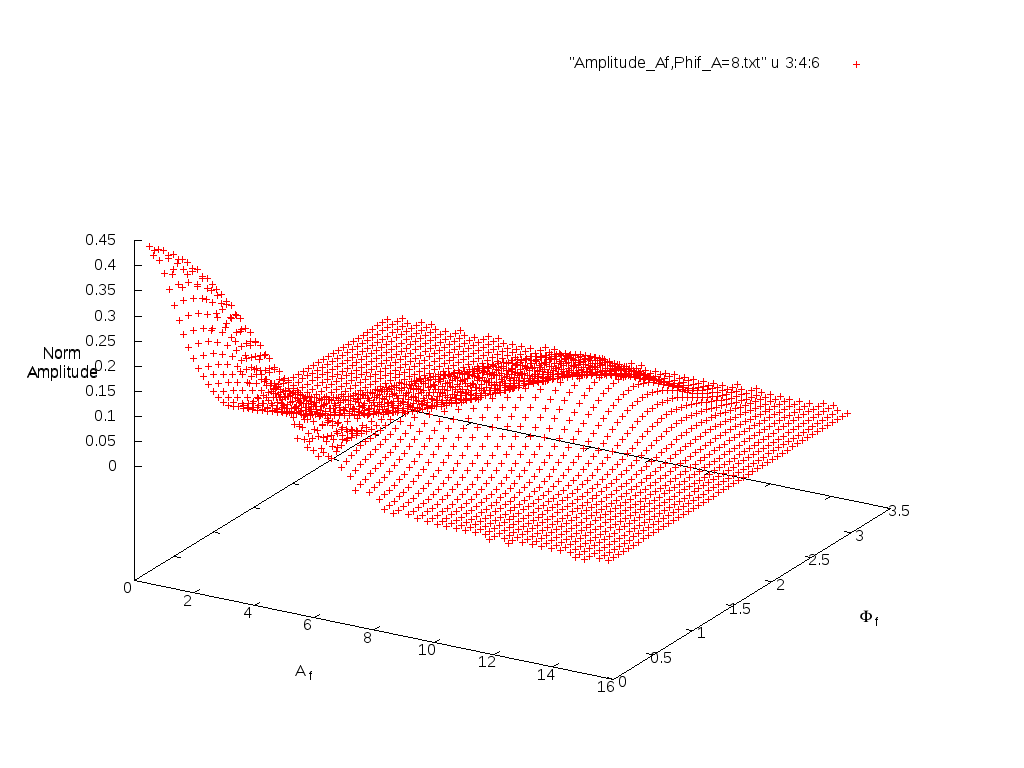}\tabularnewline
\hline 
\end{tabular}}

\centerline{\begin{tabular}{|c|c|c|}
\hline 
$A=9.28\approx\frac{2}{\sqrt{3}}j_{f}$ & $A=12.16$ & $A=15.68$\tabularnewline
\hline 
\hline 
\includegraphics[scale=0.2]{Data_Article/Amplitude_Af,Phif_A=9\lyxdot 28} & \includegraphics[scale=0.2]{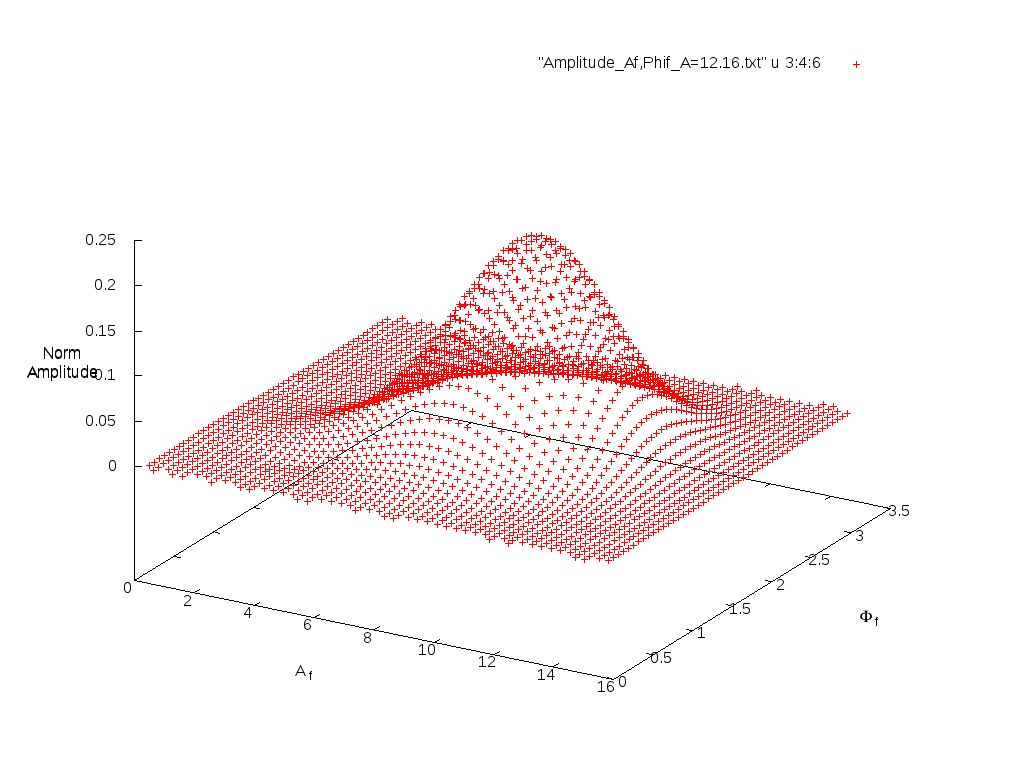} & \includegraphics[scale=0.2]{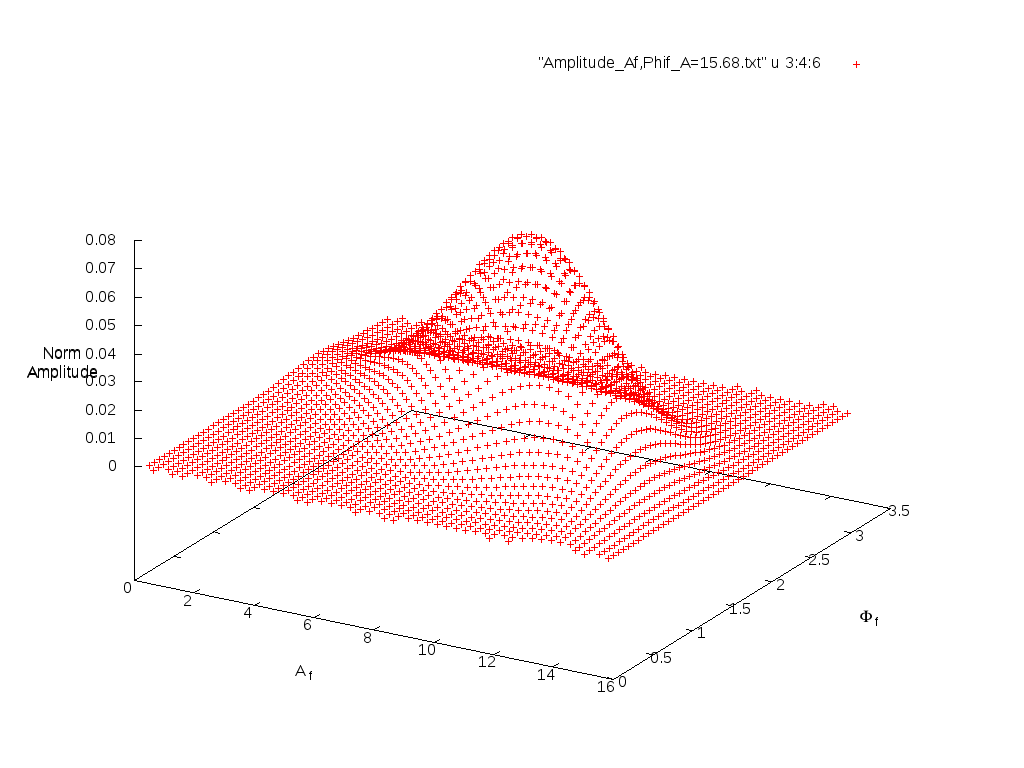}\tabularnewline
\hline 
\end{tabular}}
\caption{$f(A_f,\Phi_f,A)$, always for $j=j_{0}=j_{f}=8$.} 
\label{FG3}
\end{figure}

The last of these figures shows also that there seem to be an increase of the amplitude away from the classical values for low angles and low projected areas. To study this effect it is convenient to move away from the classical region. It is instructive to see what happens if we take a non-cassical value of the projected area $A=A_1=A_2=A_2$ of the equatorial tetrahedra. The numerical amplitude is given in Figure \ref{FG3} with different values of $A$ (the classical one is the fourth). That is, Figure \ref{FG3} plots 
\be
f(A_f,\Phi_f,A)=f_{j_{0},j,j_{f}}\left(A,A,A,A_f,A_f,\frac\pi2,\frac\pi2,\frac\pi2,\Phi_f,\Phi_f\right). 
\ee

Here we see an interesting phenomenon: there is a large peak for small areas and angles, which is not accounted for by the classical limit geometry. It is clearly an effect of degenerate geometries, as evident from the fact that it is at the angle $\Phi_f\sim 0$.  As $A$ increases and get closer to its classical value, the peak at the classica value  $\Phi_f\sim \frac\pi2$ emerges.

\subsection{Non equilateral case}

\begin{figure}[H]
\begin{center}
\begin{tabular}{|c|c|c|}
\hline 
$j_{f}=1\ (A^{classical}\approx1.15)$ & $j_{f}=4 \ (A^{classical}\approx4.61)$ & $j_{f}=5 \ (A^{classical}\approx5.77)$\tabularnewline
\hline 
\hline 
\includegraphics[scale=0.2]{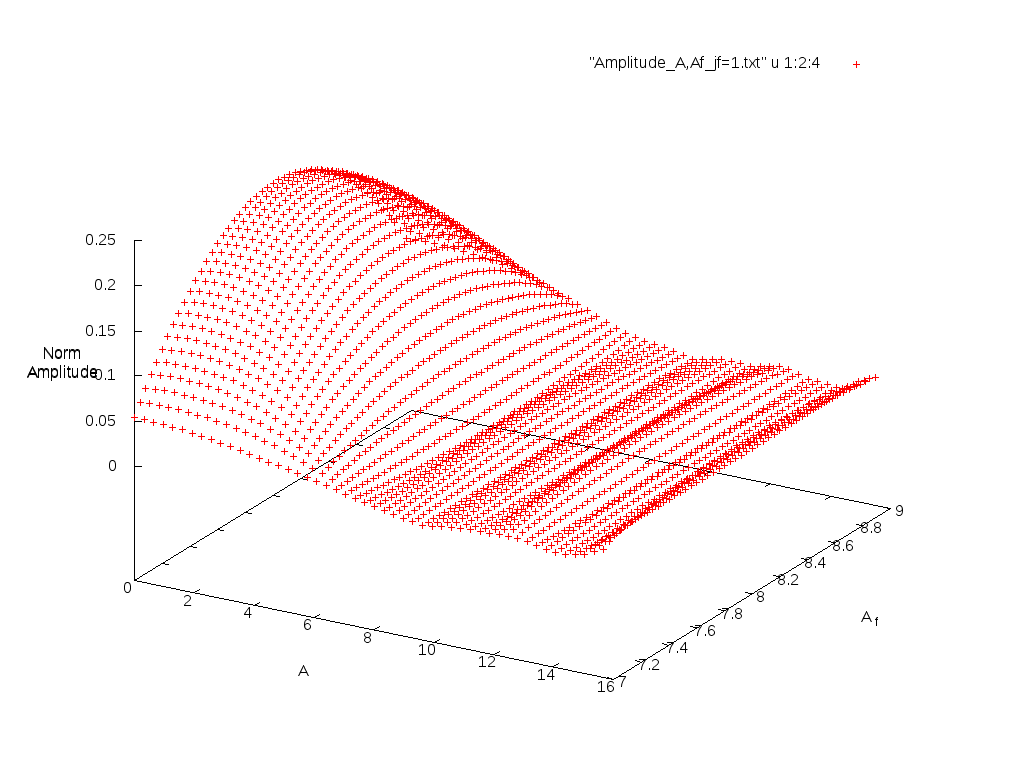} & \includegraphics[scale=0.2]{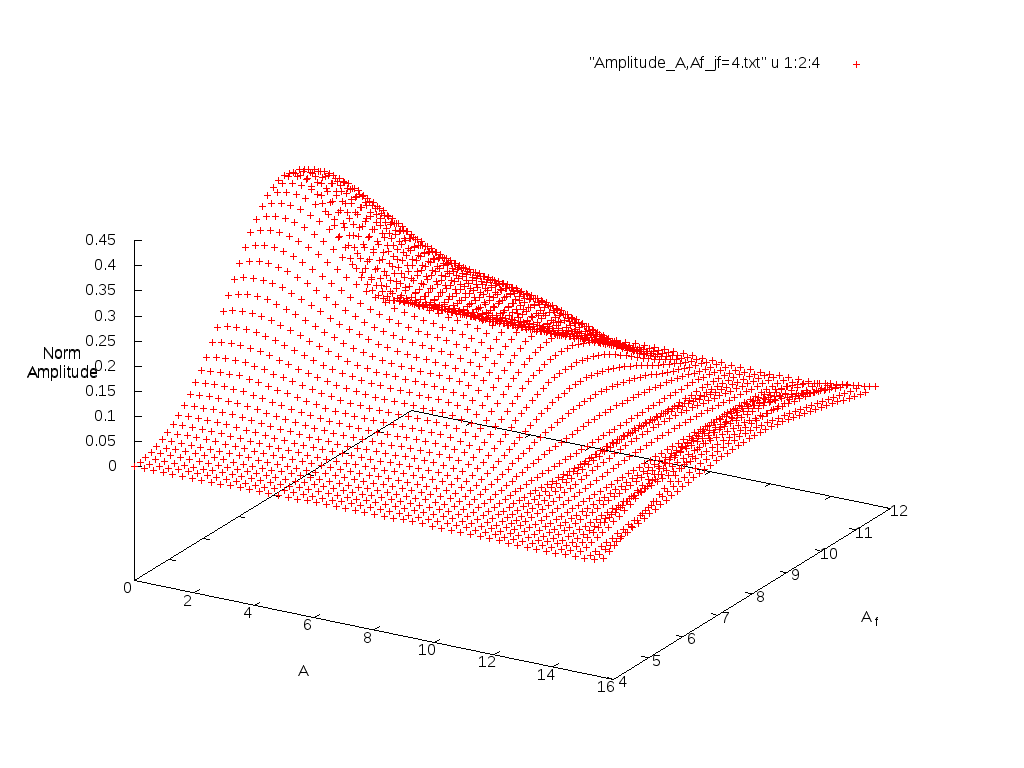} & \includegraphics[scale=0.2]{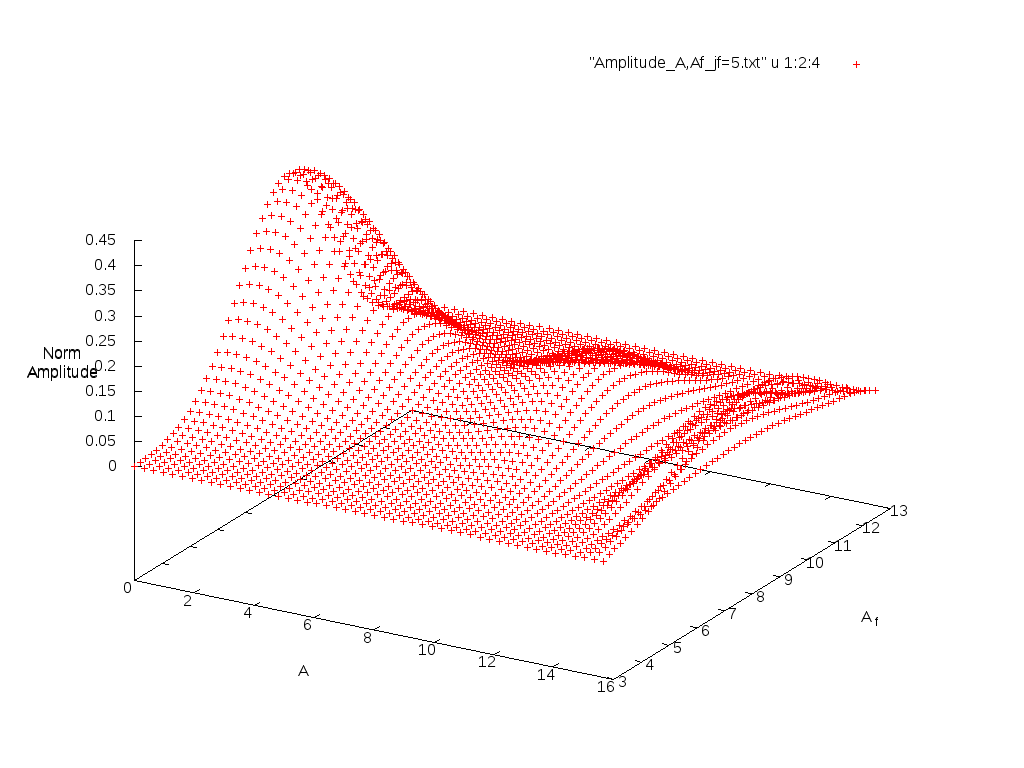}\tabularnewline
\hline 
\end{tabular}

\begin{tabular}{|c|c|c|}
\hline 
$j_{f}=7\ (A^{classical}\approx8.08)$ & $j_{f}=8 \ (A^{classical}\approx9.23)$ & $j_{f}=9 \ (A^{classical}\approx10.39)$\tabularnewline
\hline 
\hline 
\includegraphics[scale=0.2]{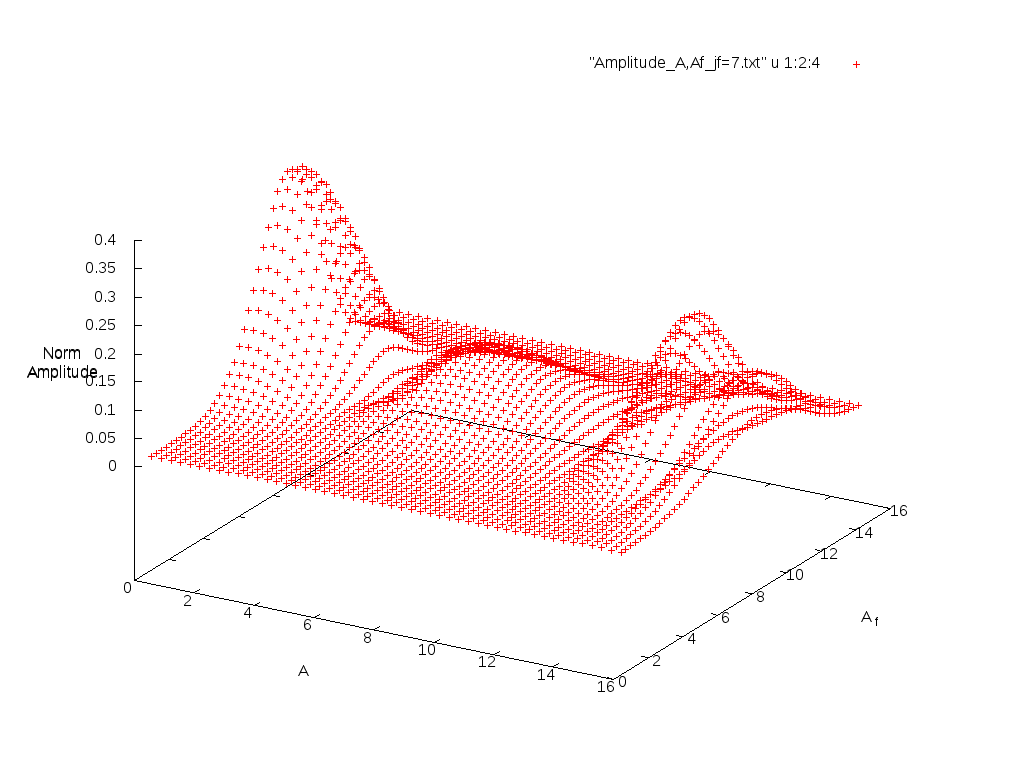} & \includegraphics[scale=0.2]{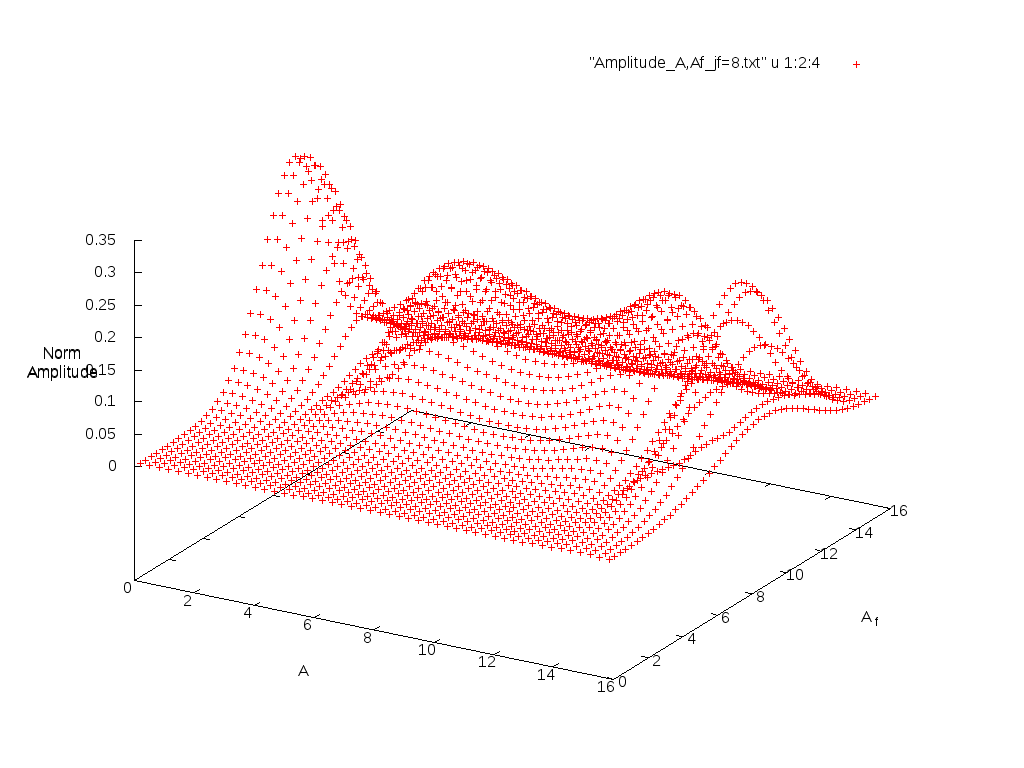} & \includegraphics[scale=0.2]{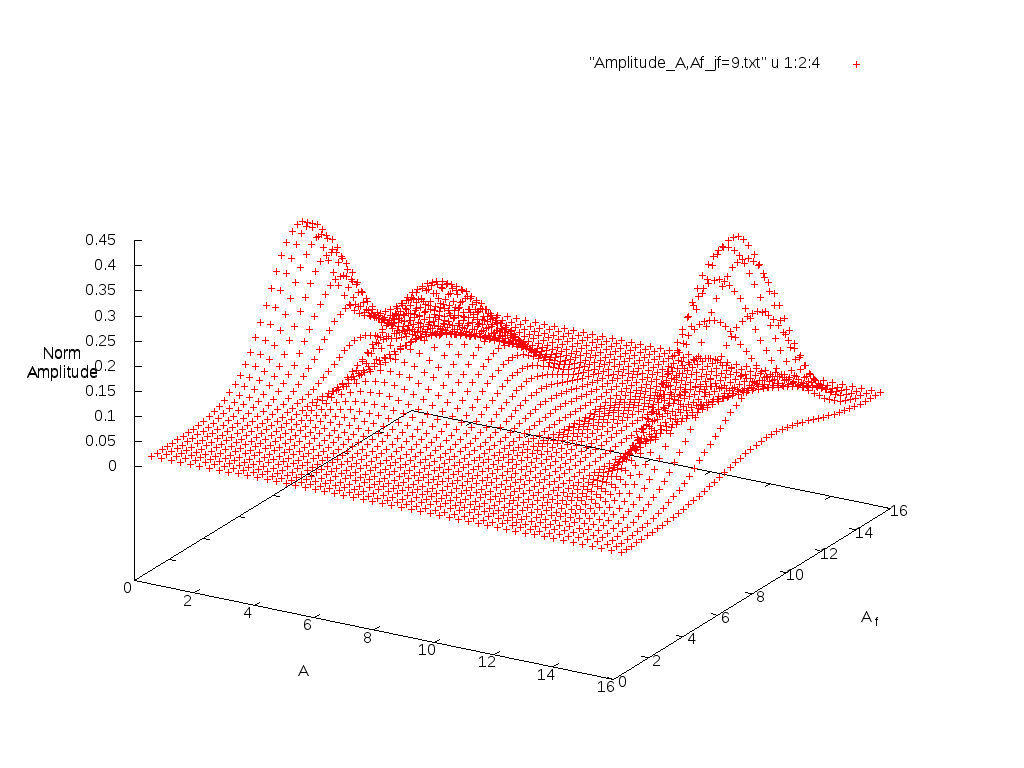}\tabularnewline
\hline 
\end{tabular}

\begin{tabular}{|c|c|c|}
\hline 
$j_{f}=11\ (A^{classical}\approx12.70)$ & $j_{f}=12 \ (A^{classical}\approx13.86)$ & $j_{f}=13 \ (A^{classical}\approx15.01)$\tabularnewline
\hline 
\hline 
\includegraphics[scale=0.2]{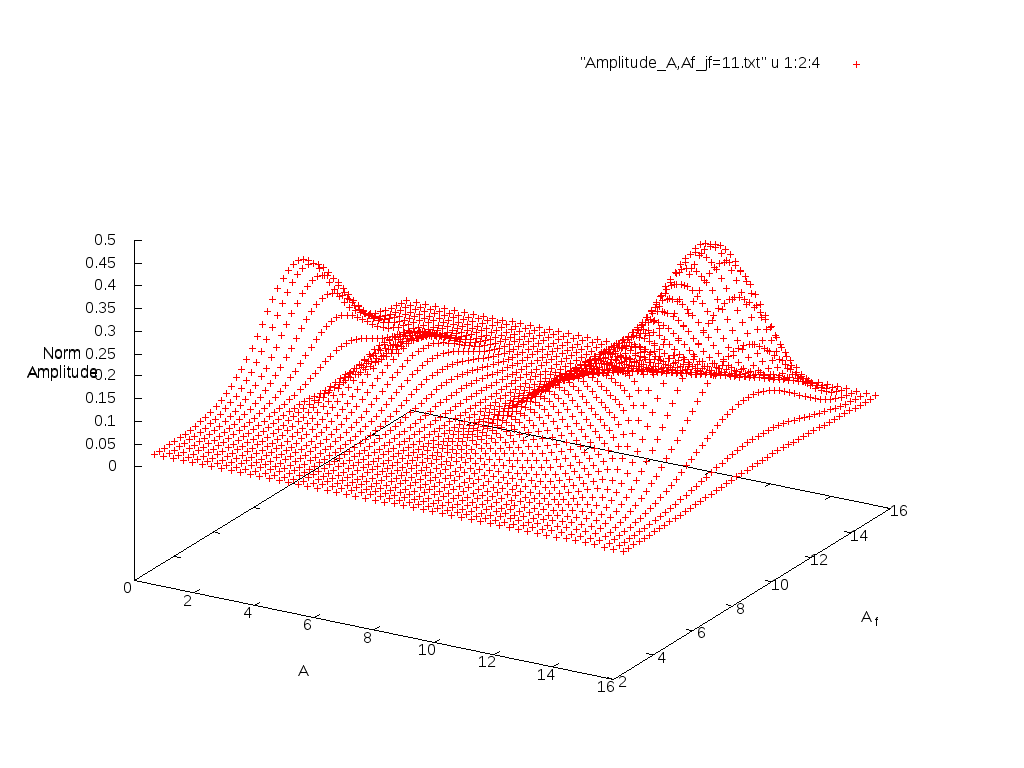} & \includegraphics[scale=0.2]{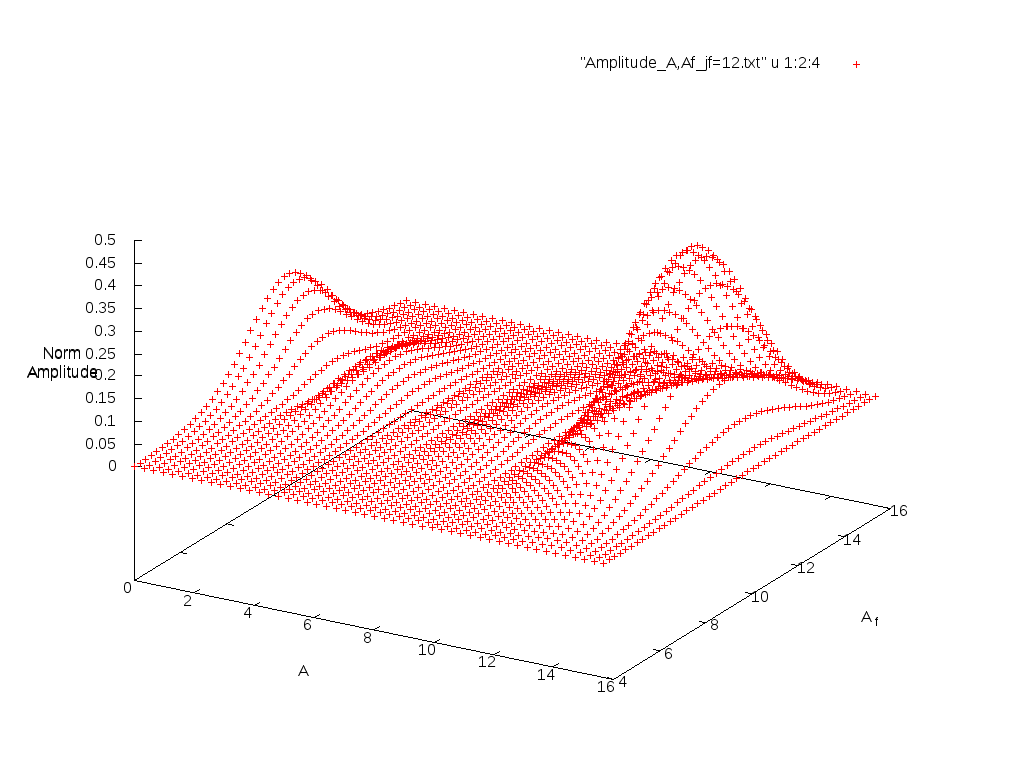} & \includegraphics[scale=0.2]{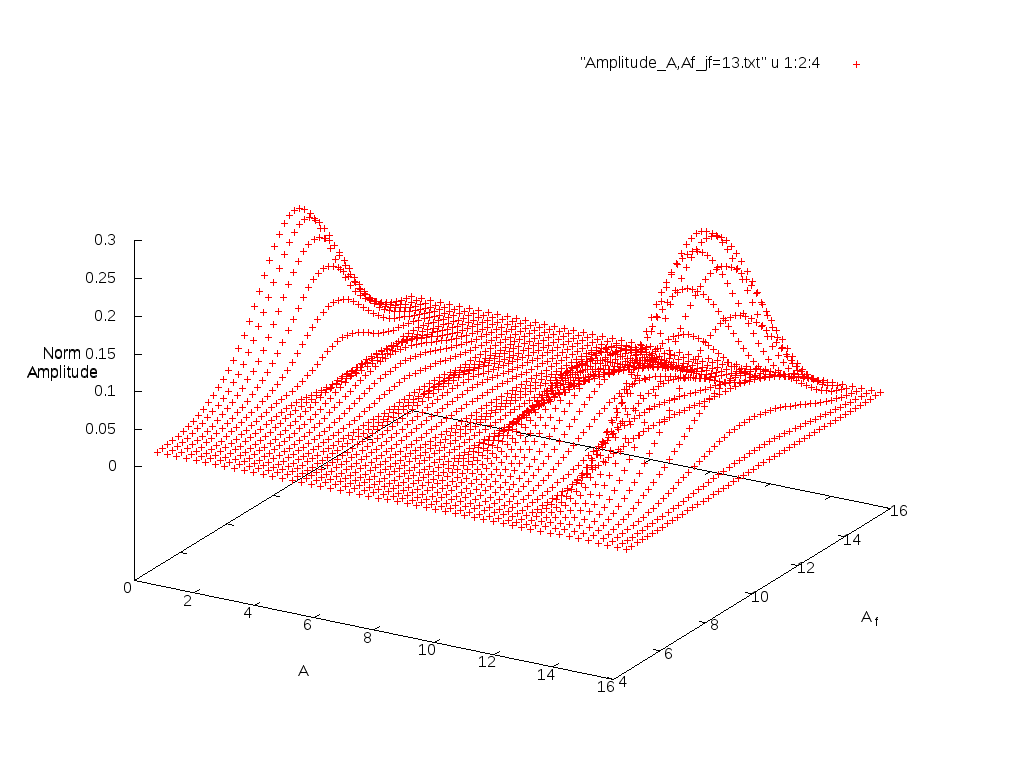}\tabularnewline
\hline 
\end{tabular}

\begin{tabular}{|c|c|c|}
\hline 
$j_{f}=15$ & $j_{f}=19$ & $j_{f}=23$\tabularnewline
\hline 
\hline 
\includegraphics[scale=0.2]{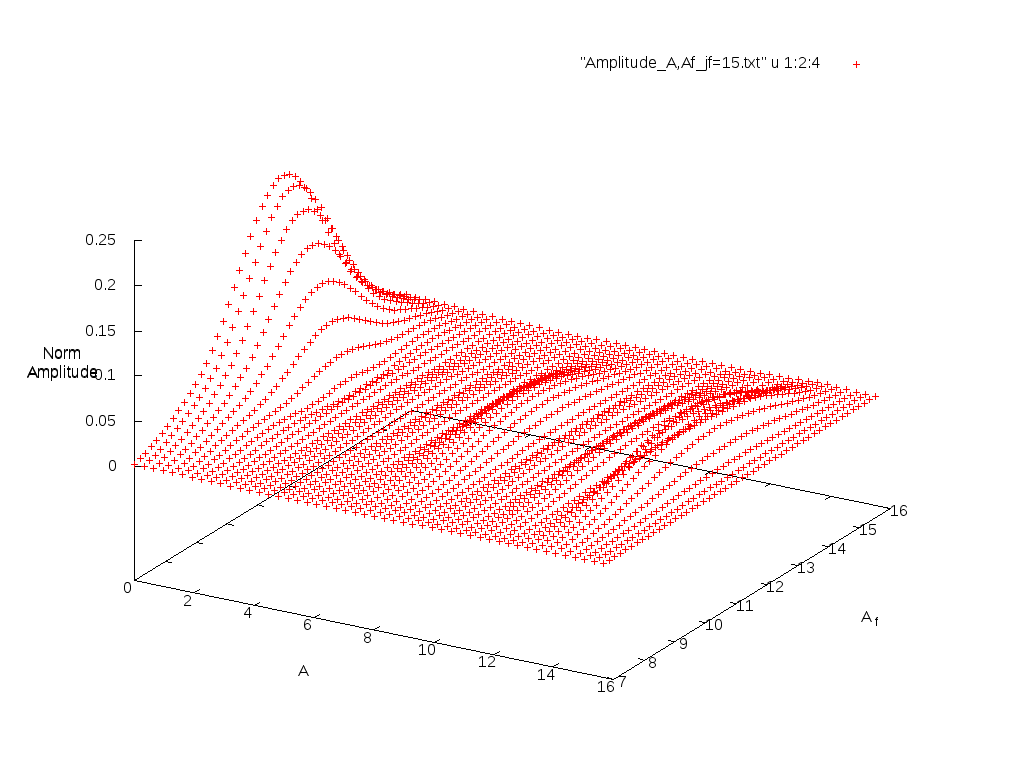} & \includegraphics[scale=0.2]{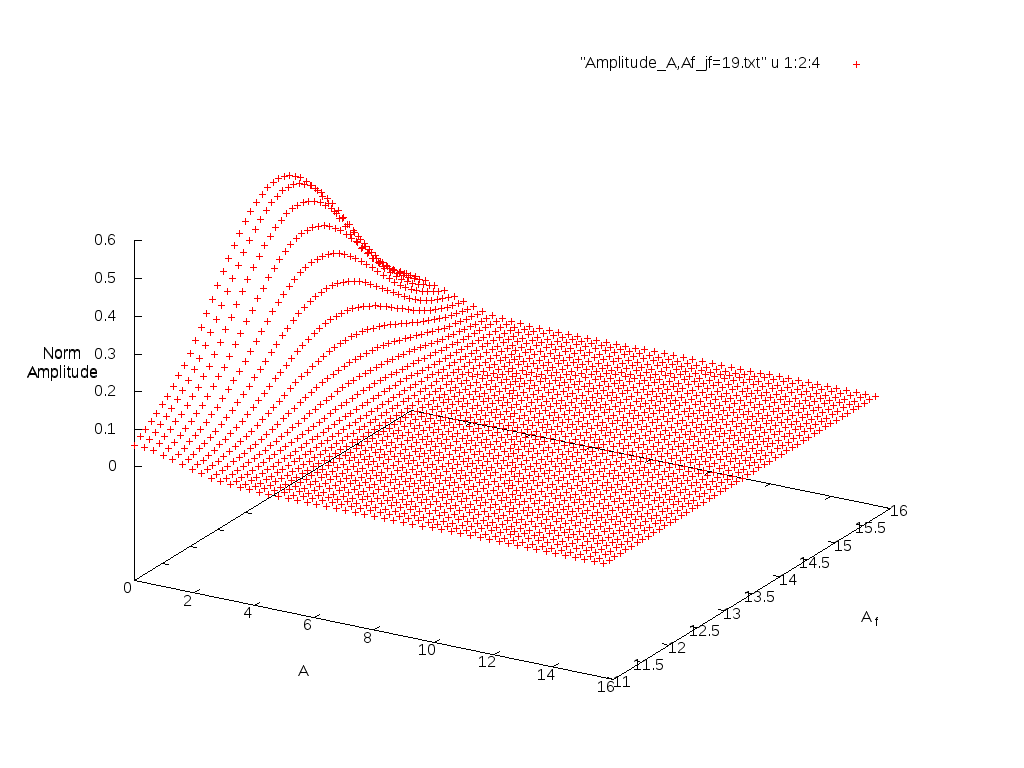} & \includegraphics[scale=0.2]{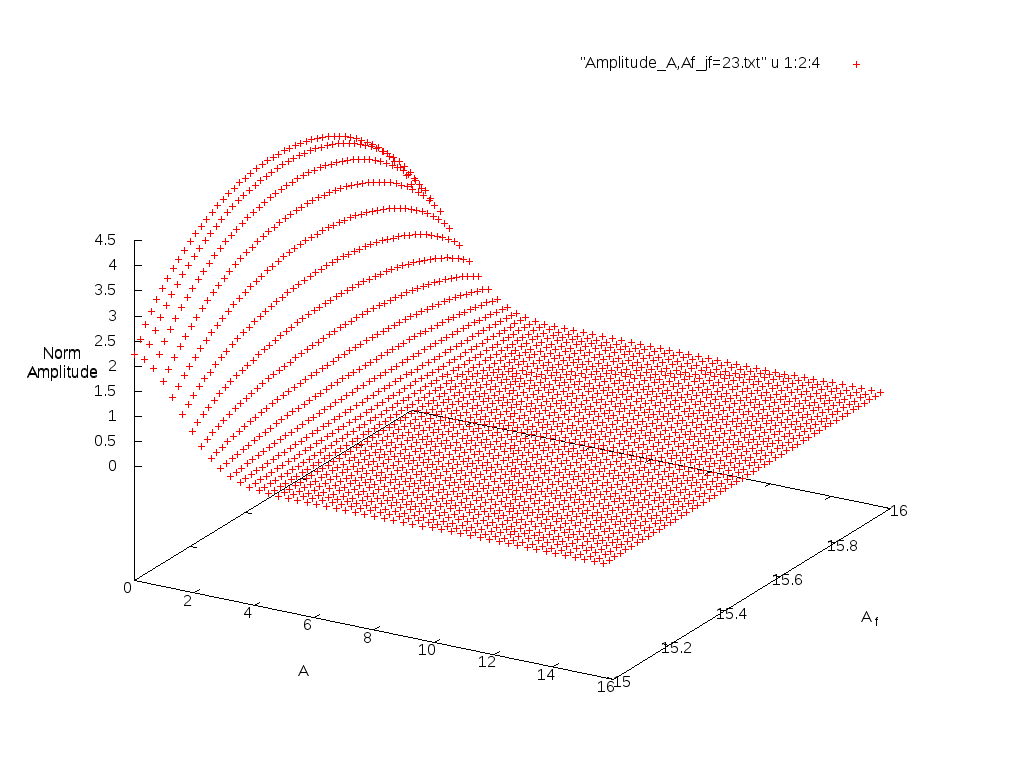}\tabularnewline
\hline 
\end{tabular}
\end{center}
\caption{$f_{j_f}(A,A_f)$}
\label{FG4}
\end{figure}

The results displayed above refer all to the equilateral case $j_{kl}=j$ which defines a regular 4-simplex. Here we give some results that move away from this case.  We keep $j=j_0=8$ and vary $j_f$.  Geometrically, $j_f$ determines the size of an equator of the 4-simplex, therefore modifies it from spherical to ellipsoidal. Figure \ref{FG4} gives the numerical evaluation of 
\be
f_{j_f}(A,A_f)=f_{j_{0}=8,j=8,j_{f}}\left(A,A,A,A_f,A_f,\frac\pi2,\frac\pi2,\frac\pi2,\frac\pi2,\frac\pi2\right). 
\ee
for different values of $j_f$. 

These results are interesting for various reasons.  First of all, they indicate that the amplitude always peaks on the correct classical value of 
$A_{f}=\sqrt{j_{0}^{2}+\frac{1}{3}j_{f}^{2}}$ given by \eqref{c1}.  However, the peak in $A$ is far less prominent. The classical constraint \eqref{c3} in this case gives
\be
      A=\frac{2}{\sqrt 3}j_f,
\ee
whose value is indicated in the Figure for each $j_f$. For small and large values of $j_f$ (compared to the $j=j_0=8$) the geometry of the triangles is nearly degenerate, and therefore degenerate configurations dominate.  This is manifest in the $j_f=1$ and $j_f=23$ cases, that descrive very ``elongated'' or very ``flattened'' 4-simplices. In these cases the quantum discreteness of the volumes prevents smooth behaviours and therefore makes the geometry fuzzy.  Furthermore, in the cases of $j_f=15$ to $j_f=23$ the classical geometry does not exist because the triangular conditions of the classical tetrahedra are not respected : the quantum cases are the only ones to exist and dominate. But also in the intermediate cases ($j_f=7$ to $j_f=11$) the classical peak is not prominent over the ``quantum noise''.  We expect the classical peak to emerge for larger spins, as dictated by Barrett's theorem, but the numerical evidence we have found for this is still weak.

\subsection{The saddle point}

\begin{figure}[H]
\centerline{\includegraphics[scale=0.5]{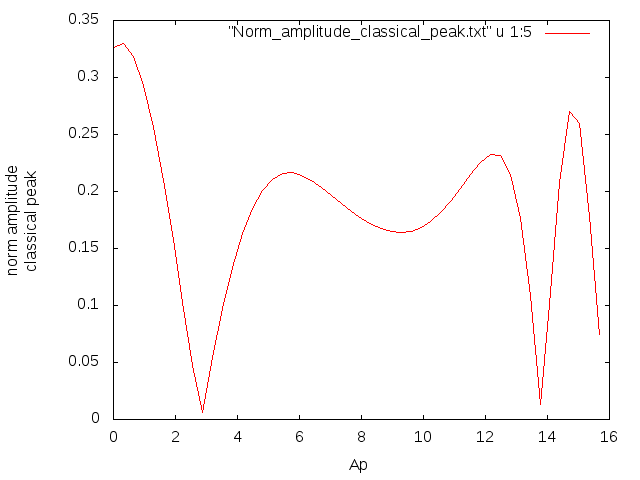}}
\caption{$f(A)$, for $j=j_{0}=j_{f}=8$.} 
\label{FG5}
\end{figure}

\begin{figure}[H]
\centerline{
\includegraphics[scale=0.3]{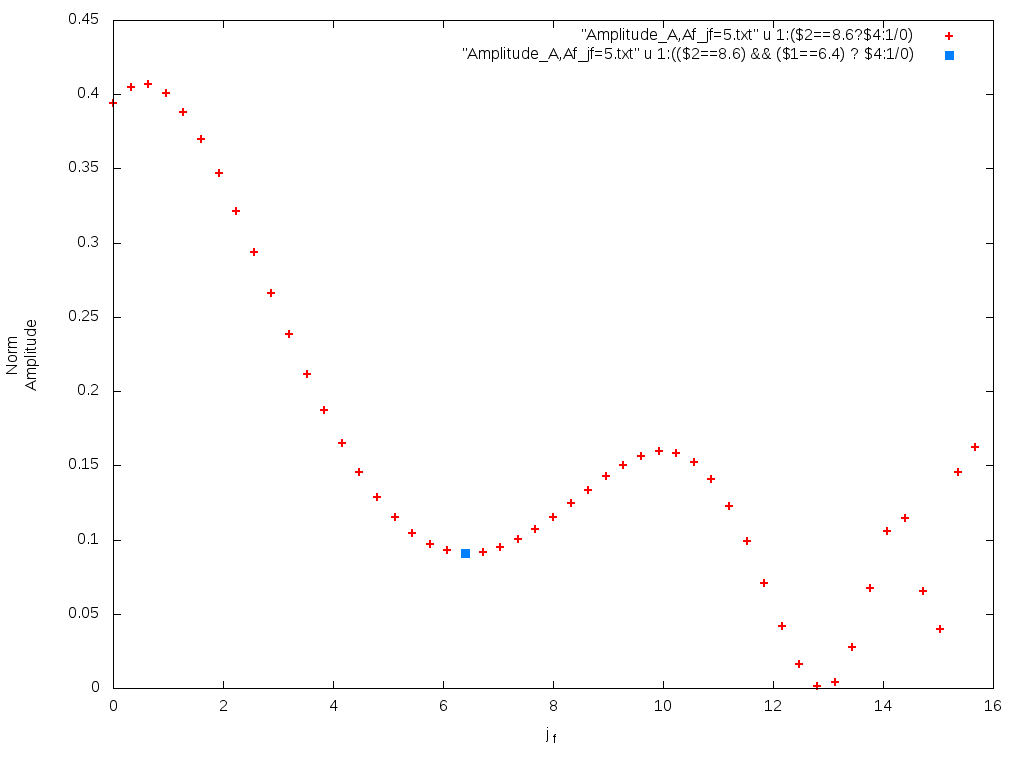}\hspace{3em}
\includegraphics[scale=0.3]{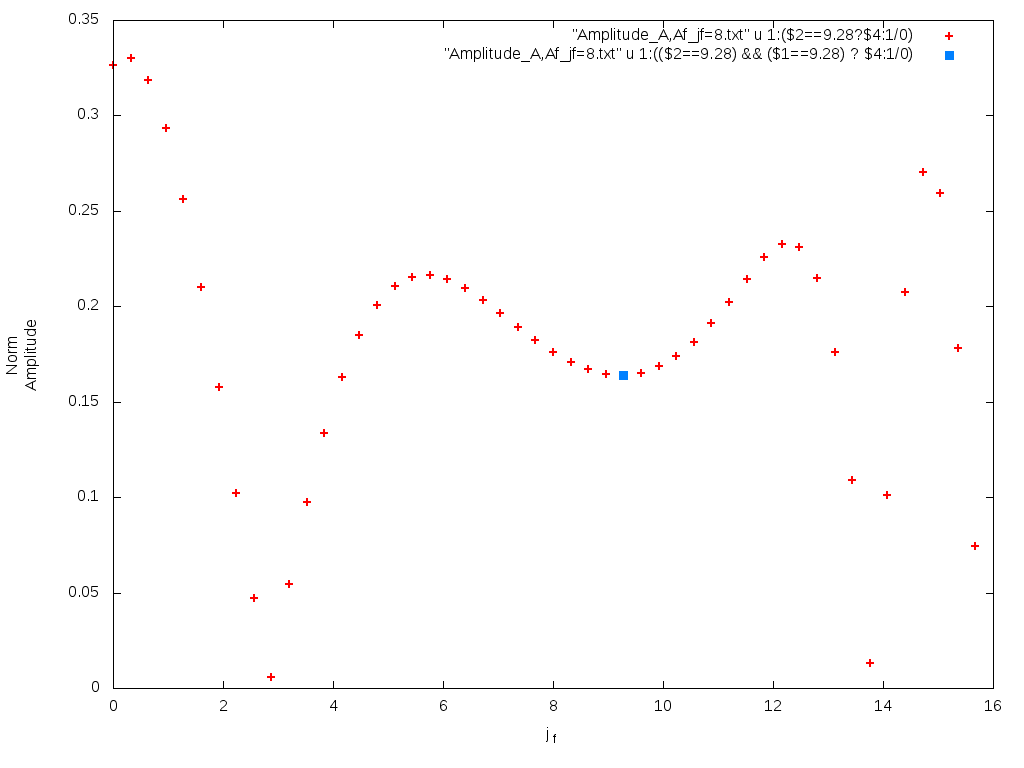}}
\caption{Left: The saddle point (in blue) for the $j_f=5$ case close to the classical value $A=2\frac{j_f}{\sqrt(3)}\approx5.77$. Right: The saddle point (in blue) for the $j_f=8$ case on the classical value $A=2\frac{j_f}{\sqrt(3)}\approx9.23$.}
\label{FG6}
\end{figure}

\begin{figure}[H]
\centerline{
\includegraphics[scale=0.3]{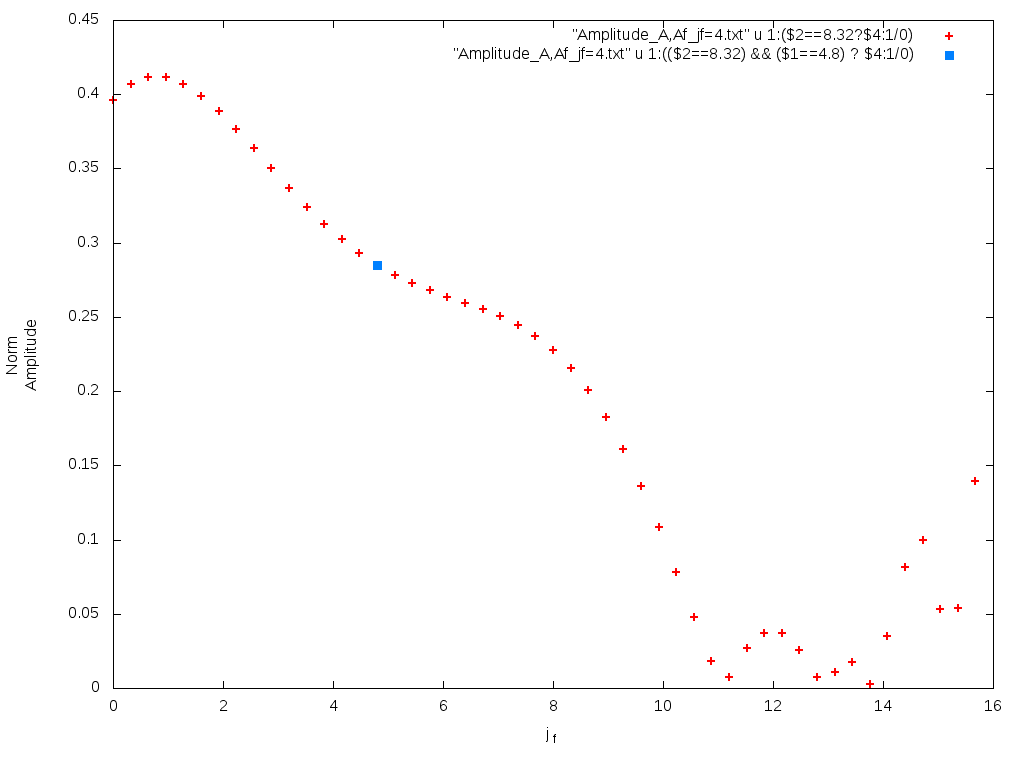}\hspace{3em}
\includegraphics[scale=0.3]{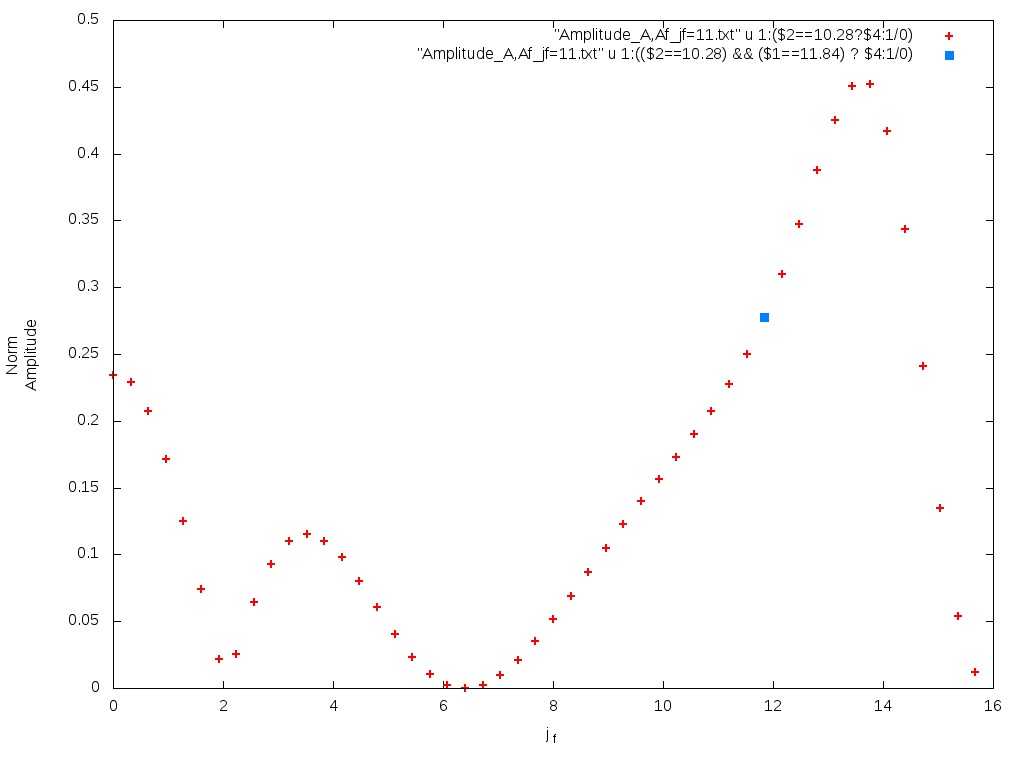}}
\caption{Left: The pseudo-saddle point (in blue) for the $j_f=4$ case on the classical value $A=2\frac{j_f}{\sqrt(3)}\approx4.62$. Right: The pseudo-saddle point (in blue) for the $j_f=11$ case close to the classical value $A=2\frac{j_f}{\sqrt(3)}\approx12.70$.}
\label{FG7}
\end{figure}

Before concluding, we point out a curious feature of the amplitude that we have found, which we have not fully understood. 
As we have seen, it is on the variable $A$ that the peakedness of the amplitude is less clear, at these spins. Back to the 
equilateral case $j_0=j=j_f$, let us plot the amplitude as a function of $A$ when the other variables are on their classical values, that is
\be
f(A)=f_{j_{0}=j=j_f=8}\left(A,A,A,A_f(j,j_0,j_f),A_f(j,j_0,j_f),\frac\pi2,\frac\pi2,\frac\pi2,\frac\pi2,\frac\pi2\right). 
\ee
The result is given in Figure \ref{FG5}.
The peaks at low and high value are given by the degenerate configurations. In the central portion of the graph, we see the classical peak emerge. However, there are two of these, and the actual classical value, which is $A=2\frac{jf}{\sqrt{3}}\approx9.28$ happen to fall on the minimum between them. Since this is still a maximum in the $A_f$ directions, it follows that this is actually a saddle point.  The same patterns seems to remain if we vary $j_f$. In Figure \ref{FG4}, in fact, for $j=5$ (away from the degenerate region) the classical value, appears to be also close to the saddle point. And if we look precisely the cases $j_f=4$ $(A^{classical}\approx4.61)$ or $j_f=11$ $(A^{classical}\approx12.70)$, where classical solutions should appear, we see the fusion of the classical and quantum regions which give a `pseudo-saddle point' on their respectives classical values. The saddle and `pseudo-saddle' points are given in Figure \ref{FG6} and \ref{FG7}.

\begin{figure}[H]
\centerline{
\includegraphics[scale=0.3]{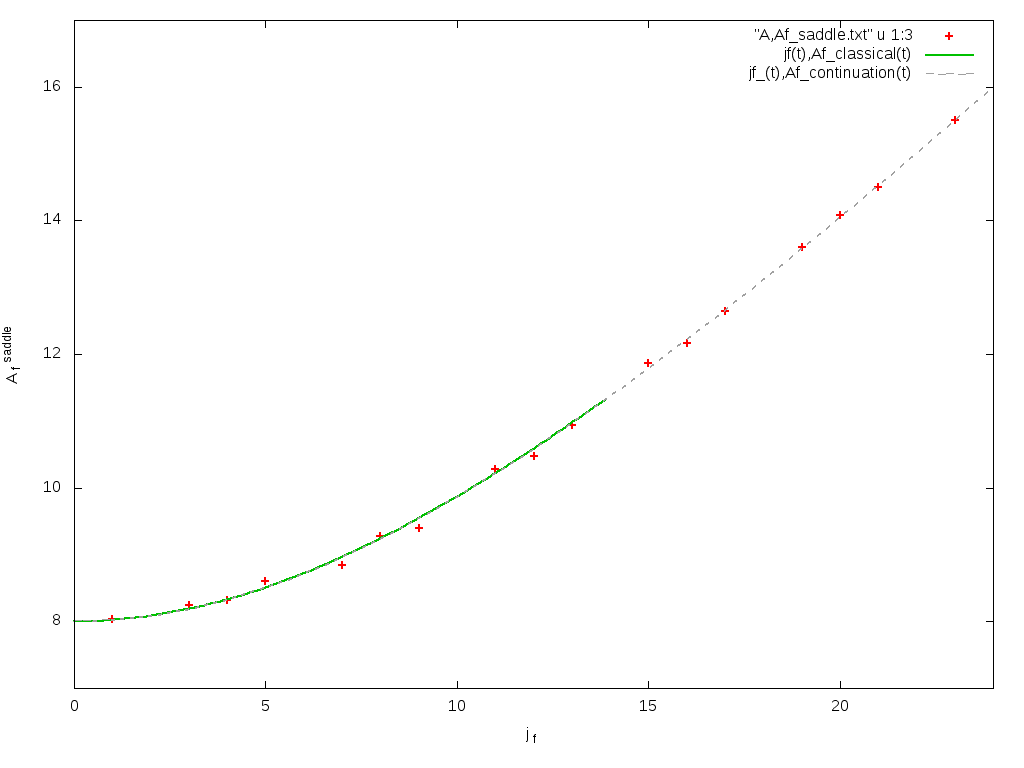}\hspace{3em}
\includegraphics[scale=0.3]{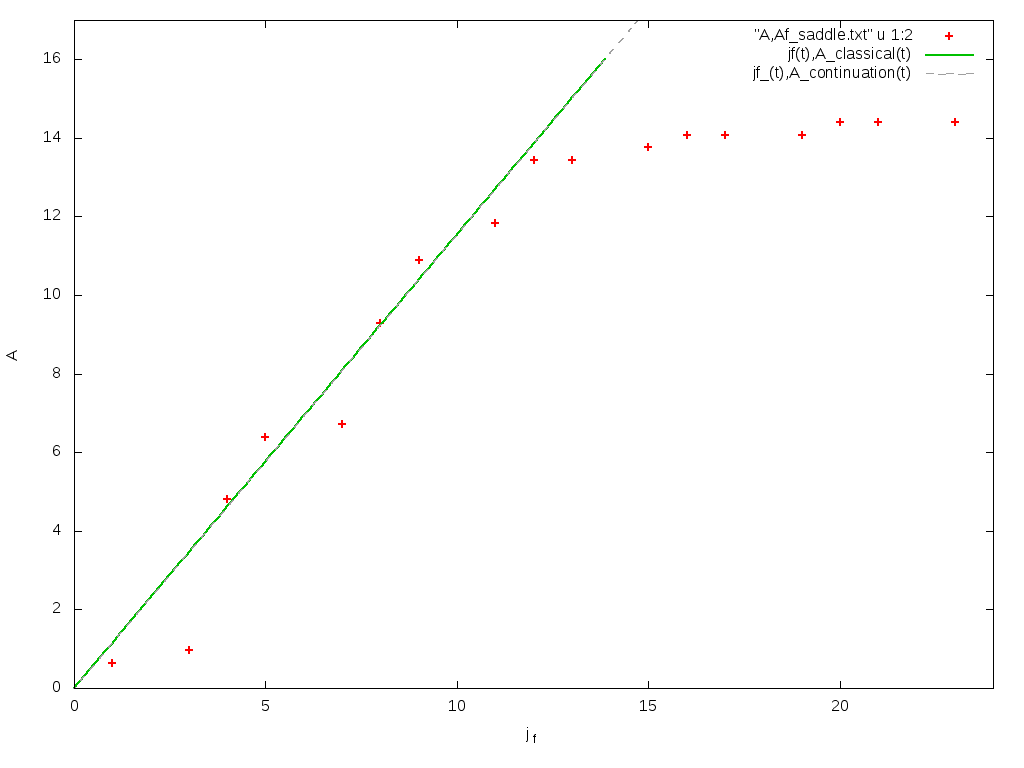} }
\caption{The position of the saddle points and peaks in  $A_f$ and $A$ as $j_f$ varies.  The green lines gives the classical values and the dotted lines are the analytic continuation of the classical values.} 
\label{FG8}
\end{figure}

Finally, we can track the saddle points and the local maximums close to the expected classical solution in function of $j_f$, Figure \ref{FG8}. 

While the saddle points and classical peaks track the classical value of $A_f$ with precision, they seems to jump from the degenerate case to some approximate values and then back to a degenerate case in the case of $A$. Overall, the global evolution of the $A$ is the same than the classical geometry, but is essentially concentrated around the degenerated regions. Also, remember these peculiar points/peaks are alway maximum in the $A_f$ direction but not in the $A$.

\begin{figure}[H]
\centerline{
\includegraphics[scale=0.3]{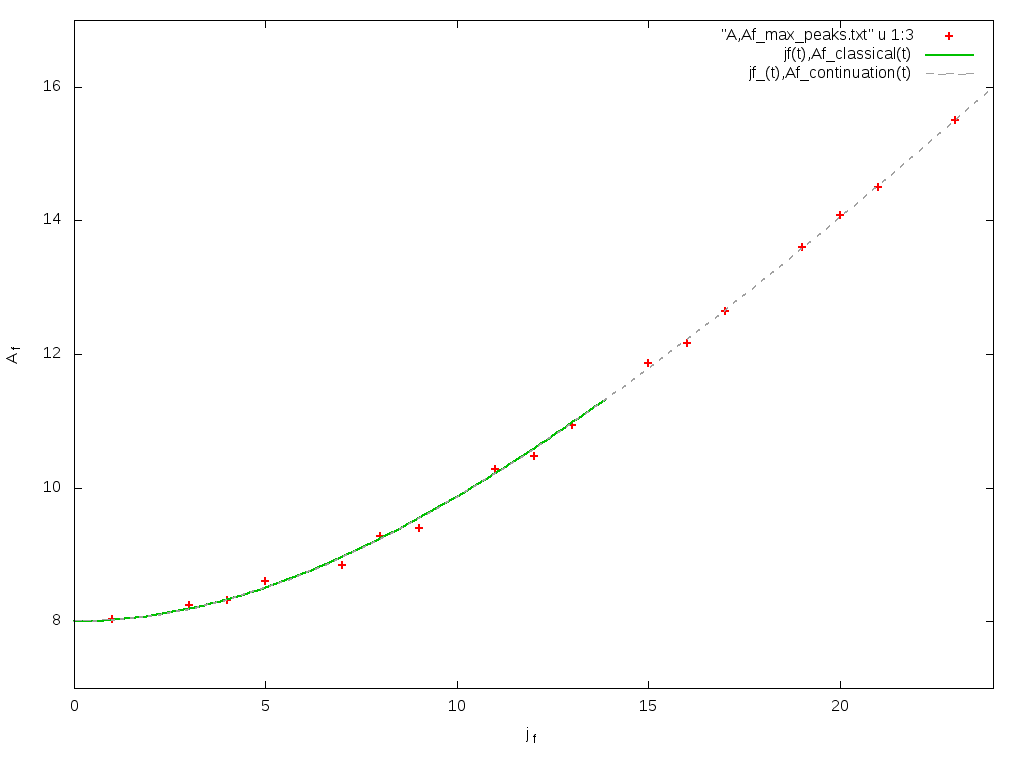}\hspace{3em}
\includegraphics[scale=0.3]{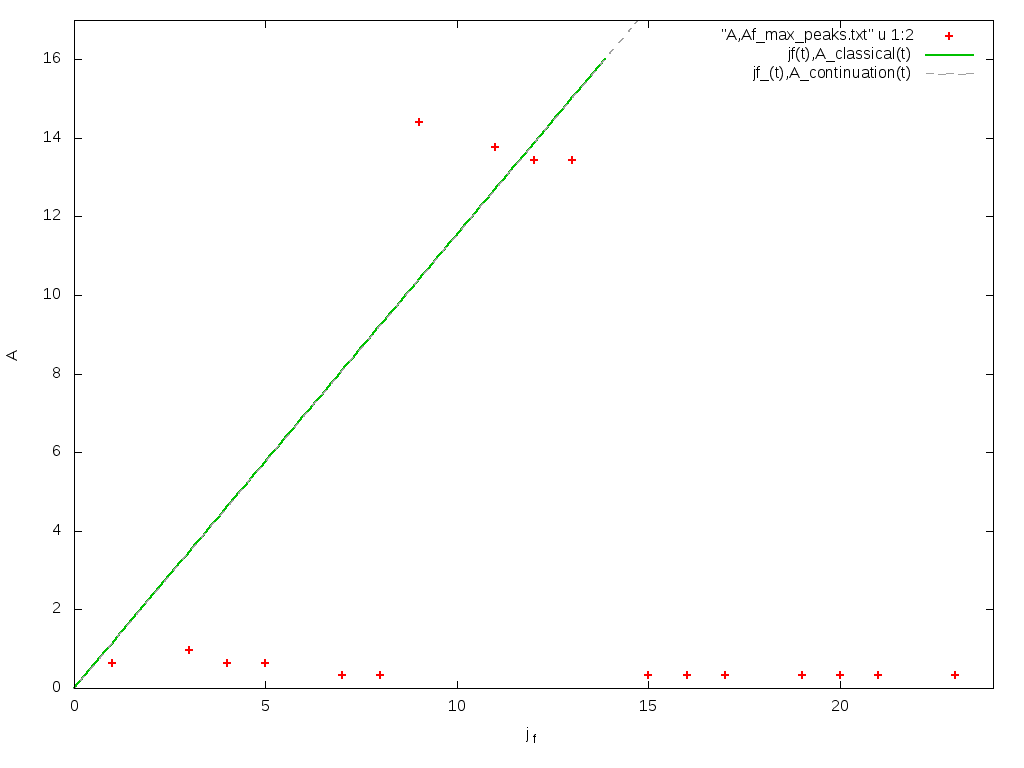} }
\caption{The position of the maximum peak in  $A_f$ and $A$ as $j_f$ varies.  The green lines gives the classical values and the dotted lines are the analytic continuation of the classical values.} 
\label{FG9}
\end{figure}

To better display the behaviour of the peaks, we have ploted the position of the highest peak of the amplitude as a function of $A_f$ and $A$, for different values of $j_f$ (always at fixed $j=j_0=8$) in Figure \ref{FG9}.
 
The peaks track again the classical value of $A_f$, they seems to jump from the degenerate case to another in the case of $A$. Even if the `saturation' in the quantum regions are understandable for the Figure \ref{FG8} and \ref{FG9} --because the classical geometry does not exist for $j_f \geq j_0\sqrt{3}$ where the triangular conditions are not respected-- we do not know how significative is this observation and whether it would permane at higher spins.

\subsection{Summary}

The classical geometry of a cylindrically symmetric 4-simplex is determined by three parameters that can be taken to be the three areas $a,a_0, a_f$. The geometry of its boundary is composed by two kinds of tetrahedra, equatorial and polar, whose shapes are described by two couples of shape parameters, which we have called  $A,\Phi$ and $A_f,\Phi_f$, respectively. When these tetrahedra bound the flat 4-simplex, these parameters take values, which we have denoted ``classical values'', determined by the three variables $a,a_0, a_f$. 

The quantum geometry of a coherent quantum state on the boundary of a single vertex is determined by the values of the areas of the triangles and the shape variables of the tetrahedra. Under cylindrical symmetry, the independent variables are again $a,a_0, a_f, A_f,\Phi_f, A,\Phi$.  In the limit of large areas, the modulus square of the amplitude is peaked on the classical values of $A_f,\Phi_f, A,\Phi$. 

Here we have studied numerically the peakedness properties of the modulus square of the amplitude for small values of the spins, up to $j\sim 10$.  We have found that the classical behaviour already emerges.  In particular, the modulus of the amplitude appear to clearly peaked on the classical values of $A_f,\Phi_f$ and $\Phi$.  The peak in $A$ is disturbed by the presence of high amplitude values around degenerate configurations for small-volume  4-simplices.  We have also observed a curious presence of two peaks and a small saddle point at the classical value of $A$, which we do not understand. 

\appendix
\section{Appendix: Tetrahedron geometry}

We express the unit normals $\overrightarrow{n_{i}}$ in polar coordinates as $\overrightarrow{n_i}=\left(\theta_{i},\phi_{i}\right)$, meaning
$\overrightarrow{n_i}=\left(\cos\phi_i\sin\theta_i,\sin\phi_i\sin\theta_i,\cos\theta_i\right)$. We 
choose the orientation of the tetrahedron by (gauge) fixing $\left(\theta_{1},\phi_{1}\right)=\left(0,0\right)$,
$\phi_{2}=0$ and $\phi_{3}\in\left[0,\pi\right]$, $\phi_{4}\in\left[\pi,2\pi\right]$. By
using these relations, straigthforward geometry gives 
\be
\begin{tabular}{ll}
$\theta_{1}=0$ & $\phi_{1}=0$\tabularnewline
$\cos\theta_{2}=\frac{A^{2}-a_{1}^{2}-a_{2}^{2}}{2a_{1}a_{2}}$ & $\phi_{2}=0$\tabularnewline
 & \tabularnewline
\multicolumn{2}{l}{$4a_{1}a_{3}A^{2}\cos\theta_{3}=\cos\Phi\sqrt{2A^{2}\left(a_{3}^{2}+a_{4}^{2}\right)-\left(a_{3}^{2}-a_{4}^{2}\right)^{2}-A^{4}}\sqrt{2A^{2}\left(a_{1}^{2}+a_{2}^{2}\right)-\left(a_{1}^{2}-a_{2}^{2}\right)^{2}-A^{4}}$}\tabularnewline
 & $-\left(A^{2}+\left(a_{3}^{2}-a_{4}^{2}\right)\right)\left(A^{2}+\left(a_{1}^{2}-a_{2}^{2}\right)\right)$\tabularnewline
 & \tabularnewline
\multicolumn{2}{l}{$4a_{1}a_{4}A^{2}\cos\theta_{4}=-\cos\Phi\sqrt{2A^{2}\left(a_{4}^{2}+a_{3}^{2}\right)-\left(a_{4}^{2}-a_{3}^{2}\right)^{2}-A^{4}}\sqrt{2A^{2}\left(a_{1}^{2}+a_{2}^{2}\right)-\left(a_{1}^{2}-a_{2}^{2}\right)^{2}-A^{4}}$}\tabularnewline
 & $-\left(A^{2}+\left(a_{4}^{2}-a_{3}^{2}\right)\right)\left(A^{2}+\left(a_{1}^{2}-a_{2}^{2}\right)\right)$\tabularnewline
 & \tabularnewline
\multicolumn{2}{l}{$\cos\phi_{3}=\frac{a_{4}^{2}\sin^{2}\theta_{4}-a_{2}^{2}\sin^{2}\theta_{2}-a_{3}^{2}\sin^{2}\theta_{3}}{2a_{2}a_{3}\sin\theta_{2}\sin\theta_{3}}$}\tabularnewline
 & \tabularnewline
\multicolumn{2}{l}{$\cos\phi_{4}=\frac{a_{3}^{2}\sin^{2}\theta_{3}-a_{2}^{2}\sin^{2}\theta_{2}-a_{4}^{2}\sin^{2}\theta_{4}}{2a_{2}a_{4}\sin\theta_{2}\sin\theta_{4}}$}\tabularnewline
\end{tabular}
\ee

In the case of cylindrical symmetry, these relations simplify.  For the equatorial tetrahedra, we find
\be
\begin{tabular}{ll}
$\theta_{1}=0$ & $\phi_{1}=0$\tabularnewline
$\cos\theta_{2}=\frac{A^{2}}{2a^{2}}-1$ & $\phi_{2}=0$\tabularnewline
$\cos\theta_{3}=-\frac{A^{2}}{4aa_{0}}$ & $\cos\phi_{3}=\frac{-A\sqrt{4a^{2}-A^{2}}}{\sqrt{16a^{2}a_{0}^{2}-A^{4}}}$\tabularnewline
$\theta_{4}=\theta_{3}$ & $\phi_{4}=2\pi-\phi_{3}$\tabularnewline
\end{tabular}
\end{equation}
For the polar tetrahedra, we find
\be\begin{tabular}{ll}
$\theta_{1}=0$ & $\phi_{1}=0$\tabularnewline
$\cos\theta_{2}=-\frac{a_{f}}{3a_{0}}$ & $\phi_{2}=0$\tabularnewline
$\theta_{3}=\theta_{4}=\theta_{2}$ & $\phi_{3}=\frac{2\pi}{3}$\tabularnewline
\multicolumn{2}{c}{$\phi_{4}=2\pi-\phi_{3}$}\tabularnewline
\end{tabular}
\ee

\bibliographystyle{utcaps}
\providecommand{\href}[2]{#2}\begingroup\raggedright\endgroup

\end{document}